\documentclass[10pt,preprint]{aastex} 

\usepackage{lscape}

\newcommand{\FeII}{Fe$\;${\small\rm II}\relax}
\newcommand{\NI}{N$\;${\small\rm I}\relax}

\newcommand{\wave}[1]{$\lambda$#1\relax}
\newcommand{\twowave}[1]{$\lambda \lambda$#1\relax}

\newcommand{\kms}{km~s$^{-1}$\relax}

\newcommand{\mucol}{$\mu$ Col}
\newcommand{\zoph}{$\zeta$ Oph}
\newcommand{\delori}{$\delta$ Ori}

\newcommand{\ghrs}{GHRS}
\newcommand{\fuse}{{\em FUSE}}
\newcommand{\copernicus}{{\em Copernicus}}
\newcommand{\hst}{{\em HST}}

\begin{document}

\title{Empirical Verification of the \ion{Fe}{2} Oscillator Strengths 
	in the {\it FUSE} Bandpass\altaffilmark{1}}

\altaffiltext{1}{Based on observations made with the NASA/ESA Hubble Space
Telescope, obtained from the data archive at the Space Telescope
Science Institute. STScI is operated by the Association of
Universities for Research in Astronomy, Inc. under the NASA contract
NAS 5-26555. }

\author{J. Christopher Howk, Kenneth R. Sembach, Katherine C. Roth, 
		\& Jeffrey W. Kruk}

\affil{Department of Physics and Astronomy\\
The Johns Hopkins University\\ Baltimore, MD, 21218}

\email{howk@pha.jhu.edu, sembach@pha.jhu.edu, 
	kroth@pha.jhu.edu, kruk@pha.jhu.edu}



\begin{abstract}

We report empirical determinations of atomic oscillator strengths, or
$f$-values, for 11 ground-state transitions of \FeII\ in the
wavelength range $1050 \la \lambda \la 1150$ \AA.  We use ultraviolet
absorption line observations of interstellar material towards stars in
the Galaxy and the Magellanic Clouds taken with \copernicus, the
Goddard High Resolution Spectrograph on-board the {\em Hubble Space
Telescope}, and the {\em Far Ultraviolet Spectroscopic Explorer}.  We
derive absolute oscillator strengths by a combination of the apparent
optical depth, component fitting, and curve-of-growth fitting
techniques.  Our derived oscillator strengths are generally in
excellent agreement with recent theoretical calculations by Raassen \&
Uylings using the orthogonal operator technique.  However, we identify
three of the eleven transitions studied here whose $f$-values seem to
be incompatible with these calculations, by as much as a factor of
two.  We suggest revisions to these $f$-values based upon our
analysis.

\end{abstract}



\section{Introduction}

The measurement and analysis of absorption lines provides the basis
for much of our understanding of the content, physical conditions, and
evolution of the gas-phase interstellar medium (ISM) in galaxies.
Measurements of gas-phase abundances using such techniques have
allowed us to study the fundamental ``cosmic'' abundances of the solar
neighborhood (e.g., Meyer, Jura, \& Cardelli 1998; Meyer, Cardelli, \&
Sofia 1997) and the composition and processing of interstellar dust in
the warm neutral medium (e.g., Howk, Savage, \& Fabian 1999;
Fitzpatrick 1997; Sembach \& Savage 1996; Sofia, Cardelli, \& Savage
1994) and the warm ionized medium (Howk \& Savage 1999).  Using
absorption line measurements from the {\em International Ultraviolet
Explorer} and the {\em Hubble Space Telescope}, quality abundance
measurements have in a few cases been extended to the ISM of the
Magellanic Clouds (Welty et al. 1999a; Roth \& Blades 1995) and to the
high velocity cloud system of the Galaxy (Wakker et al. 1999; Lu et
al. 1998).  The latter examples are potentially important as
low-redshift comparisons for the abundances derived in the damped
Lyman-$\alpha$ systems (e.g., Prochaska \& Wolfe 1999; Pettini et
al. 1997, 1999; Lu et al. 1996), which can be used to study the
chemical evolutionary history of the universe over most of a Hubble
time.

The recently-launched {\em Far Ultraviolet Spectroscopic Explorer}
(\fuse; see Moos et al. 2000; Sahnow et al. 2000), a dedicated
spectroscopic observatory operating in the wavelength range 905 to
1187 \AA, will provide a wealth of data on gas-phase abundances of
abundant elements in the local universe.  Of these, iron is
particularly important since it is both an indicator of the overall
metal-enrichment of the gas as well as a significant constituent of
interstellar dust grains (Savage \& Sembach 1996).  \fuse\ will
observe interstellar clouds towards more than 50 stars in the
Magellanic Clouds (e.g., Friedman et al. 2000), and tens of objects
projected against high-velocity clouds (e.g., Murphy et al. 2000;
Sembach et al. 2000).  \fuse\ will also provide measurements of the
abundances in low-redshift intergalactic absorbers (e.g., Oegerle et
al. 2000; Shull et al. 2000).  \fuse\ has sufficient resolution
($\lambda / \Delta \lambda \ga 15,000$) to give reliable ionic column
densities in many cases, if the adopted oscillator strengths are
reliable.

While most of the near-ultraviolet (NUV; $\lambda \ga 1200$ \AA)
ground-state transitions of \FeII\ used to determine interstellar
abundances have well-determined $f$-values that are tied to absolute
laboratory measurements (Bergeson, Mullman, \& Lawler 1994, Bergeson
et al. 1996, Mullman, Sakai, \& Lawler 1997), the far-ultraviolet
(FUV; $\lambda \la 1200$ \AA) transitions of \FeII\ are not as well
constrained.  In his new compilation of atomic oscillator strengths
for use in absorption line measurements, Morton (2000) adopts the
theoretical calculations of Raassen \& Uylings (1998; hereafter RU98)
in this wavelength range.  These calculations, performed using the
orthogonal operator technique with experimentally-determined energy
levels, provide results that agree well with the laboratory
measurement of many of the NUV transitions of both \FeII\ and
\ion{Co}{2} (with average deviations of 10\%\ or less; see RU98).

Even though the agreement between theory and laboratory measurements
for the NUV lines is encouraging, an experimental check on the
$f$-values of the commonly-observed \FeII\ transitions in the
\fuse\ bandpass is desirable.  In this work we have empirically
determined the absolute $f$-values of several FUV ground-state
transitions of \FeII.  Our emphasis is on those transitions in the
wavelength range $1100 \la \lambda \la 1150$ \AA, which contains
transitions spanning more than a factor of 30 in oscillator strength,
though we also present results for a few shorter-wavelength
transitions.  Table \ref{tab:final} summarizes the results of this
study, giving our final derived $f$-values and comparing our values
with empirical and theoretical values from the literature.  In general
we find very good agreement between our empirically-derived oscillator
strengths and the theoretical results of RU98.  There are several
exceptions, including the transitions at 1112.048, 1121.975, and
1127.098 \AA.

We have used a multi-step approach to derive the oscillator strengths
given in Table \ref{tab:final}.  First, we used observations of three
nearby stars ($\zeta$ Ophiuchi, $\delta$ Orionis, and $\mu$ Columbae)
taken with the Goddard High Resolution Spectrograph (GHRS) on-board
the {\em Hubble Space Telescope} (\hst) and literature measurements
from the
\copernicus\ satellite to determine the $f$-values of several FUV 
transitions using apparent optical depth (AOD; Savage \& Sembach
1991), component fitting (see Howk et al. 1999), and curve-of-growth
fitting methods (e.g., see Jenkins 1986).  The oscillator strength
measurements were placed on an absolute scale by reference to NUV
transitions of \FeII\ whose oscillator strengths have been measured in
the laboratory (Mullman et al. 1997; Bergeson et al. 1994, 1996).
This analysis provided reliable $f$-value determinations for \FeII\
$\lambda \lambda 1125$, 1133, 1143, and 1144\footnote{We will often
refer to transitions by rounding the Morton (2000) values for their
wavelengths downward.  Hence, \ion{Fe}{2} $\lambda1144$ refers to the
line at 1144.938 \AA.} that are in excellent agreement with the
theoretical calculations of RU98.  This portion of our study is
presented in \S
\ref{sec:ghrs}.

With these well-determined absolute $f$-values in hand, we then used
\fuse\ observations of 15 stars and curve-of-growth fitting methods to 
place the 1112, 1121, 1127, and 1142 \AA\ transitions on the same
absolute oscillator strength scale.  Most of the sightlines used for
determining these $f$-values are towards stars in the Large Magellanic
Cloud (LMC; 12 of 15 stars); the remaining stars are located in the
Small Magellanic Cloud (SMC; 2 stars) and in the low halo of the Milky
Way (1 object).  We find an oscillator strength for the 1142 \AA\
transition that agrees with the calculations of RU98, although the
other three transitions examined require adjustments of 30\% to 140\%.
This phase of the study is presented in \S \ref{sec:fuse}.

We summarize our work and discuss its implications in \S
\ref{sec:summary}.


\section{GHRS and {\em Copernicus} Measurements of \ion{Fe}{2} 
	Oscillator Strengths at Far-Ultraviolet Wavelengths}
	\label{sec:ghrs}

In this section we derive the absolute oscillator strengths of several
FUV \FeII\ transitions using archival GHRS data and literature
measurements from the \copernicus\ satellite.  

For this treatment we have chosen to focus on three well-observed,
well-studied sightlines for which high-quality GHRS data and accurate
\copernicus\ measurements exist, covering both the FUV lines 
with unknown $f$-values and the NUV lines with measured $f$-values.
The three sightlines used here are the \mucol\ (Howk et al. 1999;
Shull \& York 1977), \zoph\ (Savage, Cardelli, \& Sofia 1992; Morton
1975), and \delori\ (see Bohlin et al. 1983) sightlines.

Our reduction of the archival GHRS data is discussed in \S
\ref{subsec:ghrsreductions}.  We have used three independent
techniques to derive absolute \FeII\ oscillator strengths with these
data.  These are the three methods commonly used for deriving ionic
column densities and include the apparent optical depth method of
Savage \& Sembach (1991), which we apply to the determination of
$f$-values using the \mucol\ datasets in \S \ref{subsec:nav},
component fitting techniques, which are applied to the \mucol\
sightline in \S \ref{subsec:compfit}, and curve-of-growth fitting,
which is applied to all three sightlines in \S \ref{subsec:cog}.  We
summarize the results of our GHRS and \copernicus\ analysis in \S
\ref{subsec:ghrssummary}.

	\subsection{GHRS Observations and Data Reduction}
	\label{subsec:ghrsreductions}

We have used archival GHRS observations of \mucol\ (HD 38666),
\delori\ (HD 36486), and \zoph\ (HD 149757) to measure equivalent
widths of FUV (and some NUV) \FeII\ transitions along these three
sightlines.  Although the nominal short-wavelength limit of the GHRS
is 1150 \AA, given the low efficiency of the MgF$_2$ coatings of the
\hst\ optics at wavelengths shortward of 1150 \AA, observations at such 
wavelengths are possible for sufficiently bright targets.  Howk et
al. (1999) have demonstrated this capability, presenting GHRS
observations of \ion{Fe}{3} $\lambda1122$, the \wave{1134} triplet of
\NI, and \FeII\ $\lambda \lambda 1133$, 1143, and 1144 towards the 
low-halo star \mucol. 

A log of the GHRS observations used in this work is given in Table
\ref{tab:ghrslog}.  For the sightlines towards \mucol\ and \zoph,
details of the observations at $\lambda > 1200$ \AA\ are given by Howk
et al. (1999) and Savage et al. (1992).  The GHRS data towards
\delori\ are presented here for the first time.

Our reduction of the archival GHRS data follows that of Howk et
al. (1999).  The basic calibration makes use of the standard {\tt
CALHRS} routine\footnote{{\tt CALHRS} is part of the standard Space
Telescope Science Institute pipeline and the STSDAS IRAF reduction
package.  It is also distributed via the \ghrs\ Instrument Definition
Team for the IDL package.} using the best calibration reference files
as of the end of the GHRS mission.  The {\tt CALHRS} processing
includes conversion of raw counts to count rates and corrections for
particle radiation contamination, dark counts, known diode
nonuniformities, paired pulse events and scattered light.  The
wavelength calibration was derived from the standard calibration
tables.  The absolute wavelength scale should be accurate to $\sim
\pm1$ resolution element (see Table \ref{tab:ghrslog}).

The final data reduction was performed using software developed and
tested at the University of Wisconsin-Madison.  This includes the
merging of individual spectra and allowing for additional refinements
to the scattered light correction for echelle-mode observations.  The
inter-order scattered light removal in GHRS echelle-mode data
discussed by Cardelli et al.  (1990, 1993) is based upon extensive
pre-flight and in-orbit analysis of GHRS data and is used by the {\tt
CALHRS} routine; the coefficients derived by these authors are
appropriate for observations made through the small science aperture
(SSA).  The scattered light coefficients for the large science
aperture (LSA) observations of \mucol\ used in this work are given in
Table 2 of Howk et al. (1999).  The first-order G140M, G160M, and
G200M holographic gratings have very little scattered light, and no
adjustment has been made to the zero point of spectra taken with these
gratings.

\begin{figure}
\epsscale{0.65}
\plotone{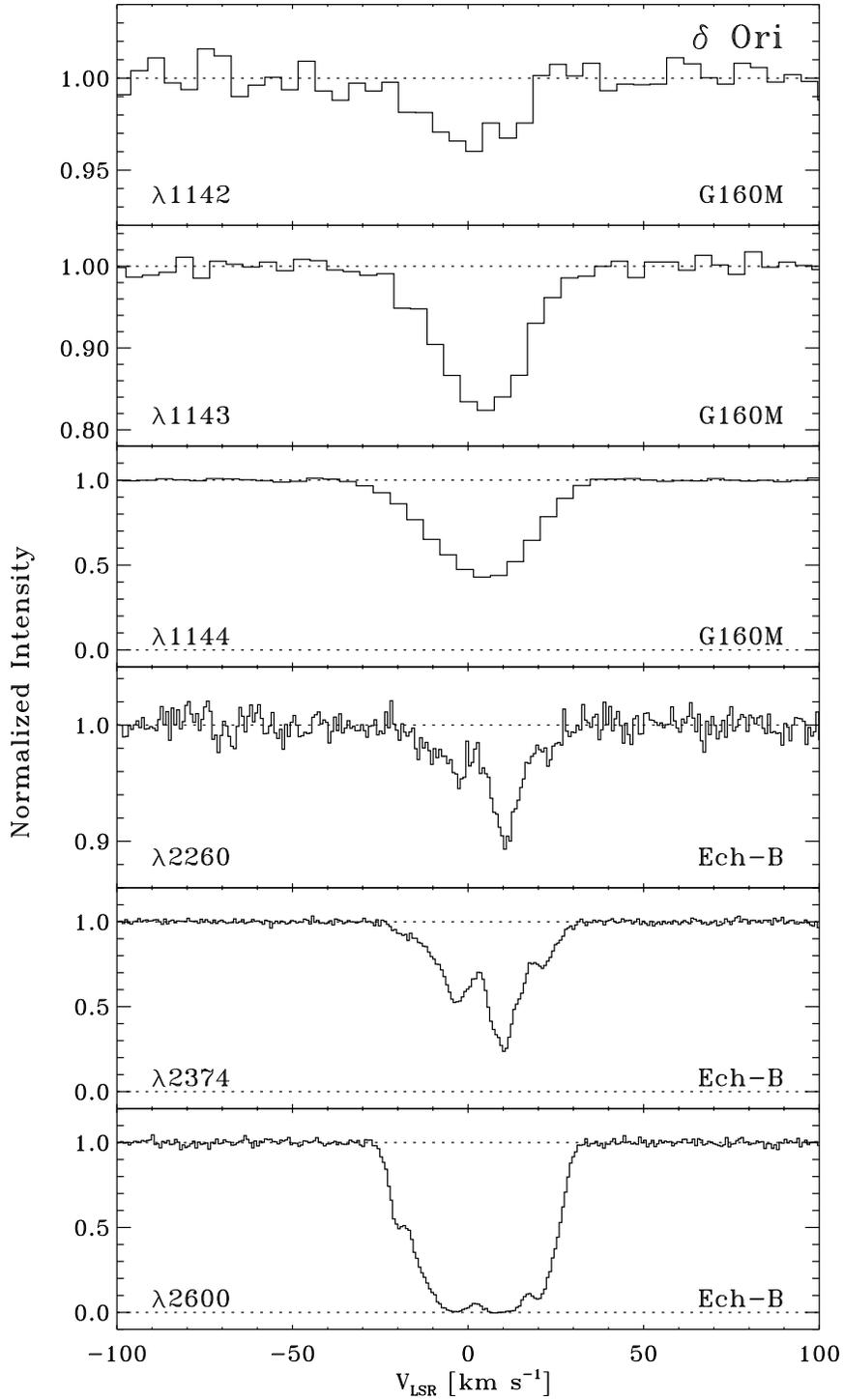}
\caption{Continuum-normalized GHRS absorption line spectra of the measured 
\protect\ion{Fe}{2} transitions towards $\delta$ Orionis.  The velocity 
scale is with respect to the local standard of rest (LSR), and no
adjustments have been made to the default wavelength calibrations of
the data.  The Ech-B data have a resolution of $\Delta v \sim 3.5$ km
s$^{-1}$; the G160M data have a resolution $\Delta v \sim 21$ km
s$^{-1}$.
\label{fig:deltori}}
\end{figure}

In all cases the GHRS absorption line data were normalized with
low-order ($<5$) Legendre polynomial fits to the local stellar
continuum, as discussed by Sembach \& Savage (1992; see their
Appendix).  Figure \ref{fig:deltori} shows the continuum-normalized
absorption profiles of the \FeII\ absorption lines towards \delori.
The top three profiles, observed with the first-order G160M grating,
have a velocity resolution of $\Delta v \sim 21$ \kms, while the lower
three lines were observed with the Ech-B echelle grating at a
resolution of $\Delta v \sim 3.5$ \kms.

We have measured the equivalent widths for interstellar \FeII\
absorption lines along these three sightlines following Sembach \&
Savage (1992); these are given in Table \ref{tab:ghrseqw}.  The error
estimates include continuum placement uncertainties and the effects of
a $2\%$ zero-level uncertainty.  Also given are \copernicus\ and GHRS
literature measurements for several transitions.  

The NUV ($\lambda > 1200$ \AA) lines listed in Table \ref{tab:ghrseqw}
are considered reference transitions in our analysis.  For these
lines, we adopt oscillator strengths derived from the quality
laboratory measurements of Mullman et al. (1997) and Bergeson et
al. (1994, 1996).  The laboratory determinations of the $f$-values for
these transitions are given in Table \ref{tab:nuvfvalues} along with
the $f$-values calculated by RU98 for comparison.

	\subsection{Apparent Optical Depth Analysis}
	\label{subsec:nav}

The apparent optical depth method for interpreting absorption line
spectra has been discussed by Savage \& Sembach (1991) and Jenkins
(1996).  Its application to empirically deriving atomic oscillator
strengths from astrophysical data has been discussed by Cardelli \&
Savage (1995) and Sofia, Fabian, \& Howk (2000).  The apparent optical
depth, $\tau_a(v)$, an instrumentally-blurred version of the true
optical depth of an absorption line, is given by
\begin{equation}
\tau_a(v) = - \ln \left[ I(v)/I_o (v) \right]
\end{equation}
where $I_o(v)$ is the estimated continuum intensity and $I(v)$ is the
observed intensity of the line as a function of velocity.  This is
related to the apparent column density per unit velocity, $N_a(v)$ [in
units ${ \rm atoms \ cm^{-2} \ (km \ s^{-1})^{-1}}$], by
\begin{equation}
N_a(v) = \frac{m_e c}{\pi e^2} \frac{\tau_a (v)}{f \lambda} = 3.768
\times 10^{14} \frac{\tau_a (v)}{f \lambda(\mbox{\AA})},
\end{equation}
where $\lambda$ is the wavelength in \AA, and $f$ is the atomic
oscillator strength.  In the absence of {\em unresolved} saturated
structure the $N_a(v)$ profile of a line is a valid,
instrumentally-blurred representation of the true column density
distribution as a function of velocity, $N(v)$.  Where unresolved
saturated structure is present, the values of the $N_a(v)$ profile are
lower limits to the true instrumentally-blurred values of $N(v)$.

For regions of absorption that are well-resolved, i.e., where $N_a(v)$
represents $N(v)$ well,
\begin{equation}
	\tau_a(v) \propto \lambda f N(v).
\end{equation}
If one compares two absorption lines from the same species, the ratio
of their apparent optical depths over this well-resolved region should
simply be the ratio of their respective values of $\lambda f$.  Hence,
the apparent column density profiles of multiple \FeII\ transitions
can be used to derive relative oscillator strengths.

The sightline towards \mucol\ exhibits a blend of warm components
centered on $v_{\rm LSR} \approx +3$ \kms\ (component 1 of Howk et
al. 1999) stretching from $v_{\rm LSR} \approx -17$ to $+16$ \kms.
This blend of components is well-resolved with the echelle-mode
gratings of the GHRS.  Howk et al. (1999) have shown lines of \FeII\
as strong as $\lambda2374.461$ ($f=0.0313$) and $\lambda2586.650$
($f=0.0691$) exhibit no unresolved saturation over this velocity range
(see their Figure 9).  Components 2 through 4 ($v_{\rm LSR} \approx
+16$ to +47) along this sightline show varying degrees of unresolved
saturated structure in the strongest lines.

We have used the $N_a(v)$ profiles of \FeII\ $\lambda \lambda1143$ and
1144 derived from GHRS echelle-mode observations of the \mucol\
sightline to determine the $f$-values of these transitions by
comparison with the $N_a(v)$ profiles of several NUV lines of \FeII.
We have limited our use of the $N_a(v)$ profiles for determining
$f$-values to those FUV transitions that were observed with the
echelle-mode gratings on the GHRS with high signal-to-noise.  This
restriction is adopted to avoid problems with unresolved saturated
structure.  This limits our use of the $N_a(v)$ method of deriving
oscillator strengths to the observations of \FeII\ $\lambda
\lambda1143$ and 1144 towards \mucol.

Using the NUV \FeII\ 2249.877, 2374.461, and 1608.451 \AA\ profiles as
reference transitions, we have calculated the value $f_\lambda /
f_{ref}$ that minimizes $\chi^2$ for each pair of unknown (FUV) and
reference (NUV) line profiles.  We have used only absorption from
component 1 ($v_{\rm LSR} \approx -17$ to $+16$ \kms) to avoid
potential unresolved saturated structure.  The presence of such
structure in the strong $\lambda 1144$ transition would reveal itself
through a divergence of the $N_a(v)$ profiles for highest $\tau_a(v)$
of the strong line compared with a weaker transition.  We find no
evidence for such saturation in the $\lambda 1144$ profile compared
with, e.g., the $\lambda 1608$ profile.  However, the noise in the
$\lambda 1144$ profile can cause points with high $\tau_a(v)$ to have
artificially high optical depths.  To avoid biases associated with low
signal to noise over a limited velocity range, we have restricted our
analysis to data points having $\tau_a \leq 2.5$ in determining the
$f$-values using the $\lambda 1144$ profile.

Table \ref{tab:ghrs} summarizes all of our GHRS $f$-value
determinations, giving the RU98 theoretical $f$-value, the average of
our determinations for each line, the individual $f$-value
measurements, and the star, instrument, and method used in deriving
those measurements.  The oscillator strengths derived as described
above are marked AOD (apparent optical depth) in Table \ref{tab:ghrs}
and represent the average value of the $f$-values derived using each
of the three reference transitions.  For \twowave{1143 and 1144} we
derive $f = 0.0206(8)$ and $0.107(4)$, respectively, using the
apparent optical depth method.  The errors include contributions from
the uncertainties in the reference oscillator strengths, as well as
the sources of error discussed by Howk et al. (1999).

\begin{figure}
\epsscale{0.99}
\plotone{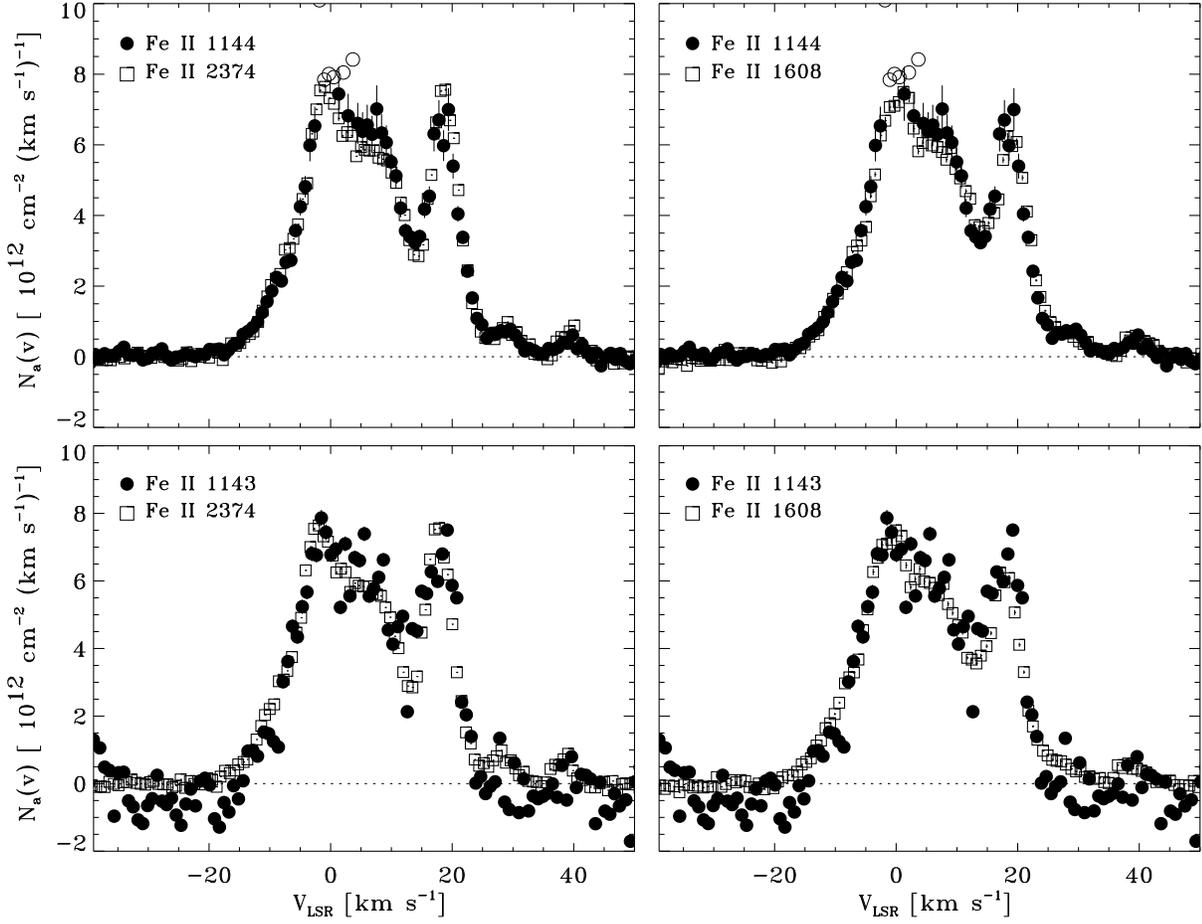}
\caption{Apparent column density profiles for the FUV transitions of 
\protect\ion{Fe}{2} at 1143.226 and 1144.938 \protect\AA\ compared with the 
reference lines at 1608.451 and 2374.461 \protect\AA\ for the sightline to
$\mu$ Col.  The circles show the FUV profiles while the open squares
denote the reference lines.  The open circles in the plots containing
the $\lambda1144$ profile show points that were not used in
determining the $f$-values because $\tau_a \geq 2.5$.
\label{fig:nav}}
\end{figure}

The $N_a(v)$ profiles for the FUV \FeII\ lines \twowave{1144 and 1143}
are compared with the reference transitions \twowave{1608 and 2374} in
Figure \ref{fig:nav} using our final $f$-values from Table
\ref{tab:final}.  The profiles of the reference lines are plotted using open
squares, while the circles show the \twowave{1144 and 1143} profiles.
For the lines shown in Figure \ref{fig:nav}, the agreement between the
$N_a(v)$ profiles is excellent across the entire velocity range.

	\subsection{Component Fitting Analysis} \label{subsec:compfit}

The component structure of the neutral ISM along the \mucol\ sightline
has been studied through an analysis of nearly 50 lines of 14 species
from dominant ionization states (Howk et al. 1999).  We make use of
the Howk et al. component structure to determine the $f$-values of the
FUV \FeII\ lines, following Fitzpatrick (1997) and Sofia et
al. (2000).  This method of analysis is most reliable when using
high-resolution data, as the details of the component structure along
a sightline can be hidden at lower resolution.  Hence we do not apply
this method to the FUV intermediate-resolution GHRS data for any of
the sightlines, which precludes use of the \delori\ and \zoph\
sightlines for determining $f$-values in this way.

We use the component-fitting software originally written by
E. Fitzpatrick and described by Spitzer \& Fitzpatrick (1993) and
Fitzpatrick \& Spitzer (1997).  This code has been updated to account
for the post-COSTAR instrumental line spread function of GHRS
observations made through the LSA (see Appendix A of Howk et
al. 1999).  A model for the interstellar medium along the sightline
was constructed by assuming a number of ``clouds,'' or components,
along the line of sight.  Each component was defined by its column
density, central velocity, and Dopper spread parameter ($b$-value).
These three parameters were varied for each component until a minimum
$\chi^2$ value between the model, convolved by the instrumental spread
function, and the original data was found.  In the case of \FeII,
which has many ultraviolet transitions, the component structure was
tested simultaneously against all of the available absorption profiles
(each of which was given an input $f$-value).

We have derived oscillator strengths of the FUV 1143 and 1144 \AA\
transitions by simultaneously fitting all of the \FeII\ lines (see
Table \ref{tab:ghrseqw} and Howk et al. 1999) observed with the
echelle-mode observations, allowing not only the parameters of the
component model to vary, but also the $f$-values of the FUV
transitions.  The final derived component structure is
indistinguishable from the results presented in Tables 5 and 6 of Howk
et al. (1999), which is expected given that the same NUV transitions
are used here to define the component structure while allowing the FUV
$f$-values to vary in the fit.  Table \ref{tab:ghrs} gives the
resulting $f$-values for $\lambda \lambda 1143$ and 1144 with
$\pm1\sigma$ error estimates.  We have allowed for a $4\%$ zero-point
uncertainty in the FUV lines.  This seems prudent given that the
observations were made through the LSA at wavelengths for which
relatively little scattered light information exists for the GHRS
echelle modes.

\begin{figure}
\epsscale{0.65}
\plotone{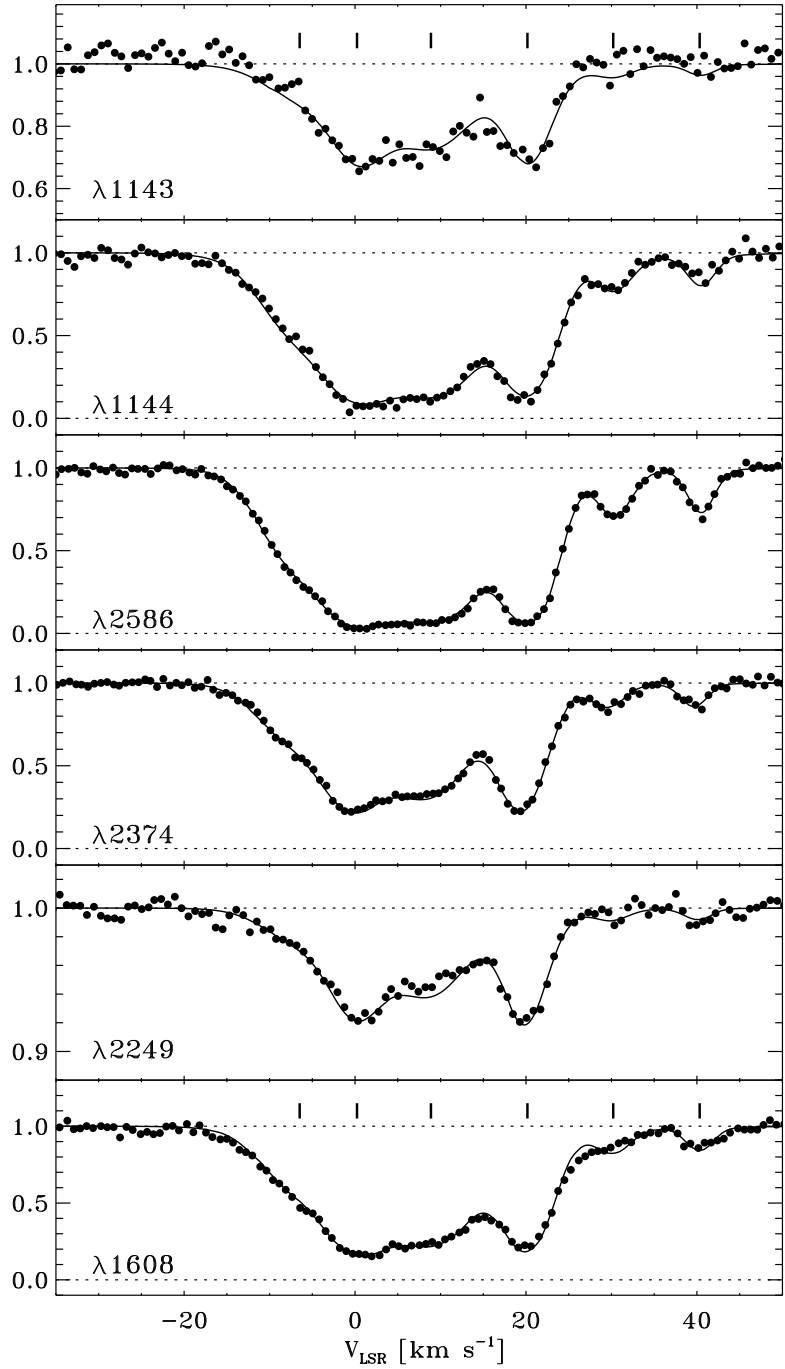}
\caption{Continuum-normalized GHRS absorption line spectra of several 
\protect\ion{Fe}{2} transitions towards $\mu$ Col.  The 
filled dots show the GHRS data, while the solid lines show the
best-fit component model of the absorption along this sightline.  The
ticks above the $\lambda \lambda 1608$ and 1143 profiles show the
locations of the velocity centroids for the adopted component model.
The model profiles use the $f$-values from Table
\protect\ref{tab:nuvfvalues}, except for the  $\lambda
\lambda 1143$ and 1144 models.  For these lines we use the best-fit 
$f$-values determined through our component fitting analysis (see
Table \protect\ref{tab:final}).
\label{fig:compfit}}
\end{figure}

Figure \ref{fig:compfit} shows the observed absorption profiles of
several \FeII\ profiles from the \mucol\ dataset with the component
model overplotted.  We have assumed the best-fit $f$-values derived
through this method when displaying the 1143 and 1144 \AA\ model
profiles, namely $f = 0.0181$ and 0.120, respectively.  The ticks
above the \twowave{1143 and 1608} profiles show the locations of the
centroids of the components in the assumed model (see Howk et
al. 1999).

	\subsection{Curve of Growth Fitting Analysis}
	\label{subsec:cog}

The ionic column densities along a sightline can be derived by fitting
a curve of growth to the measured equivalent widths (Jenkins 1986;
Spitzer 1978).  Most often this approach assumes the absorption along
a sightline can be approximated by a single Gaussian (Maxwellian)
absorption model (e.g., Morton 1975).  While this approximation is not
always valid and can yield erroneous results, particularly when using
lines of large peak optical depths (see, e.g., Jenkins 1987), it is
possible to derive accurate column densities through this method
(Jenkins 1986).

We use curve of growth fitting for the \mucol, \zoph, and \delori\
sightlines, using the equivalent widths given in Table
\ref{tab:ghrseqw}, to derive oscillator strengths of the FUV \FeII\
$f$-values.  To do this, we fit a single-component Maxwellian curve of
growth to the NUV lines of \FeII\ for each star.  Then, using the
derived column densities and $b$-values, we determine the oscillator
strength required to make the measured equivalent widths for each FUV
line in Table \ref{tab:ghrseqw} lie on the derived curve of growth.
Recent studies have used similar curve of growth fitting techniques to
derive oscillator strengths of \ion{C}{1} (Zsarg\'{o}, Federman, \&
Cardelli 1997), \ion{Ni}{2} (Zsarg\'{o} \& Federman 1998), and
\ion{Fe}{2} (Cardelli \& Savage 1995).

\begin{figure}
\epsscale{1.0}
\plotone{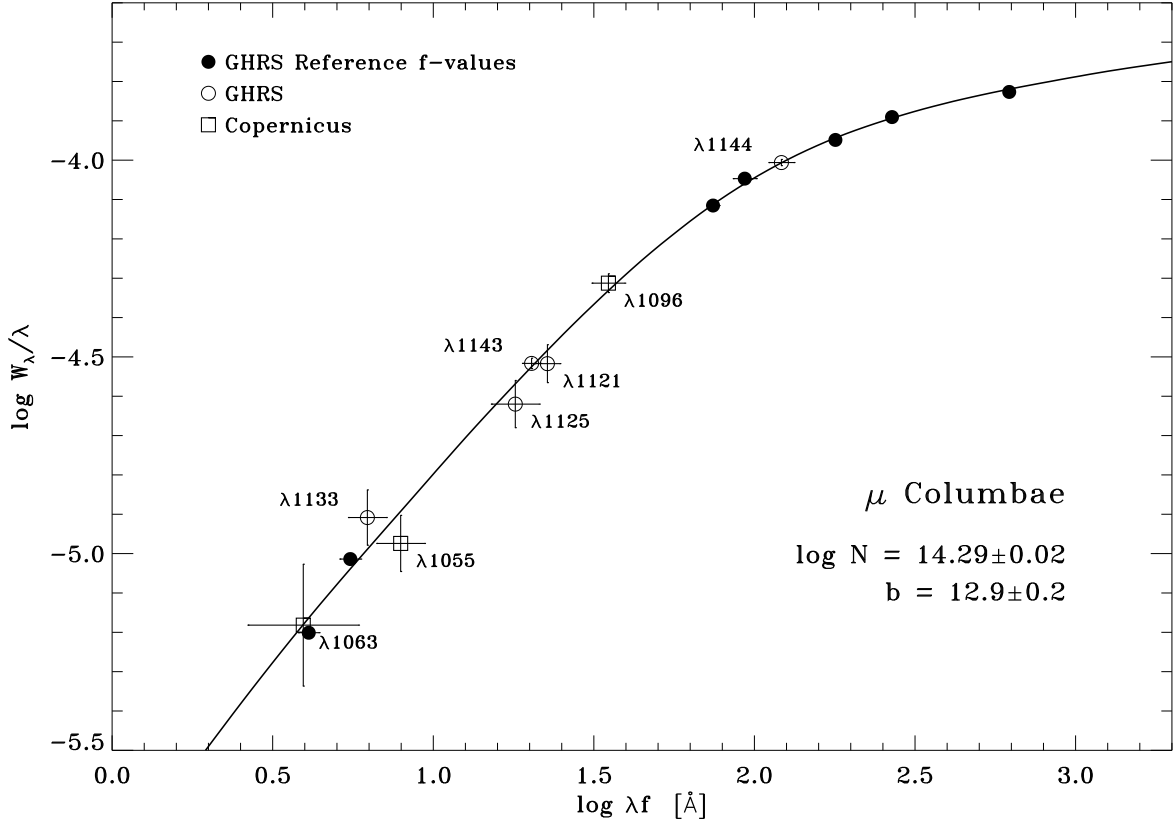}
\caption{The best-fit, single-component curve of growth for the $\mu$ Col
\protect\ion{Fe}{2} data.  This curve was fit to GHRS measurements of 
transitions with well-determined $f$-values (filled circles).  The
GHRS measurements (open circles) and {\em Copernicus} measurements
(open squares) of FUV transitions are placed on this diagram using our
final oscillator strength estimates (see Table
\protect\ref{tab:final}).  The error bars in the equivalent width
measurements for the reference transitions are typically smaller than
the filled plotting symbols.  The column density derived from this
curve of growth, $\log N(\mbox{\ion{Fe}{2}}) = 14.29\pm0.02$, is
consistent with the value $\log N(\mbox{\ion{Fe}{2}}) = 14.31\pm0.01$
adopted by Howk et al. (1999) for this sightline using
component-fitting techniques.
\label{fig:mucolcog}}
\end{figure}

The results of this fitting process are summarized in Table
\ref{tab:ghrs}.  Figure \ref{fig:mucolcog} shows the curve of growth 
fit to the measured equivalent widths for the \mucol\ sightline.  The
solid points show the NUV lines used to derive the curve of growth,
while the open symbols show the FUV measurements assuming the final
$f$-values summarized in Table \ref{tab:final}.  The derived
sightline-integrated column density, $\log N(\mbox{\FeII}) =
14.29\pm0.02$, is consistent with the value adopted by Howk et
al. (1999), $\log N(\mbox{\FeII}) = 14.31\pm0.01$, based upon
component fitting and apparent optical depth techniques.  The best fit
$b$-value is $b=12.9\pm0.2$ \kms.

\begin{figure}
\epsscale{1.0}
\plotone{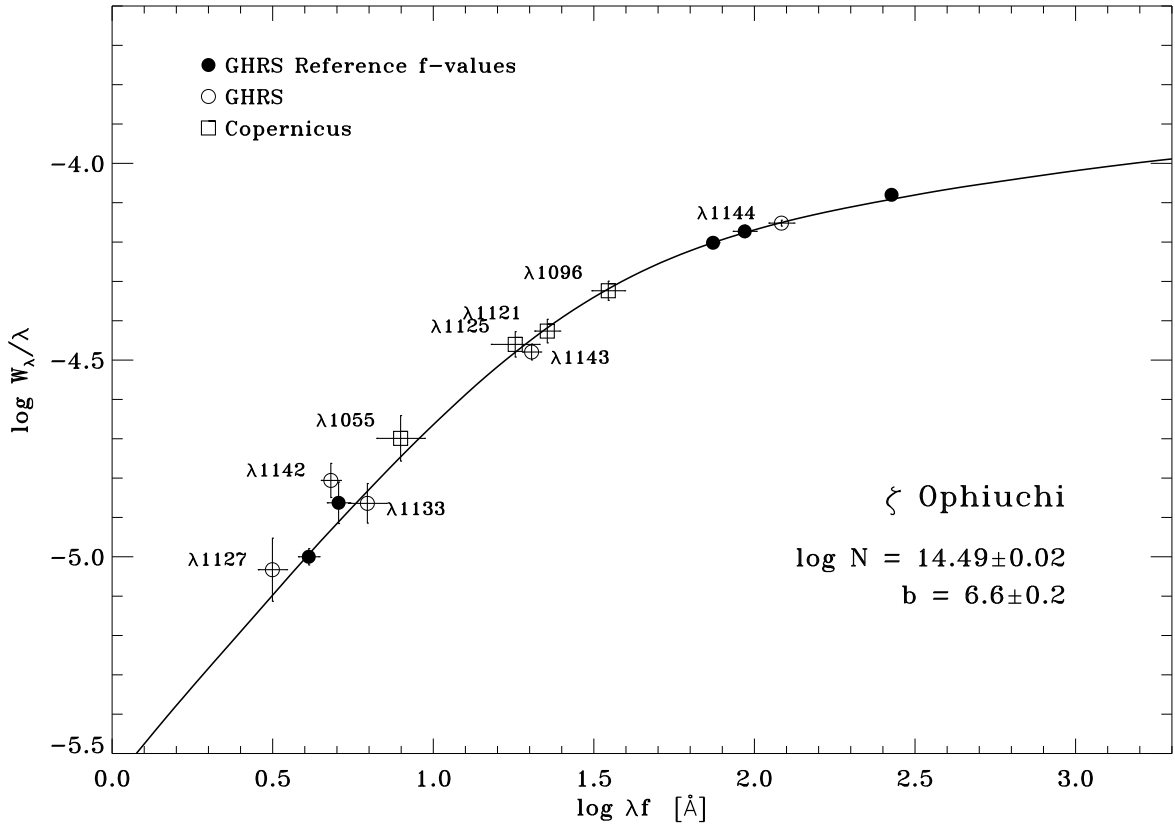}
\caption{As Figure \protect\ref{fig:mucolcog}, but for the sightline 
towards $\zeta$~Oph.  The best fit column density, $\log
N(\mbox{\ion{Fe}{2}}) = 14.49\pm0.02$, is consistent with the value
$\log N(\mbox{\ion{Fe}{2}}) = 14.51\pm0.02$ adopted by Savage \&
Sembach (1996).
\label{fig:zophcog}}
\end{figure}

For the \zoph\ sightline we derive $\log N(\mbox{\FeII}) =
14.49\pm0.02$; Savage \& Sembach (1996) derive $\log N(\mbox{\FeII}) =
14.51\pm0.02$.  The best fit $b$-value is $b=6.6\pm0.2$ \kms,
consistent with the \copernicus\ study of Morton (1975).  This curve
of growth is shown in Figure \ref{fig:zophcog}.  The FUV lines are
placed on this diagram using the final $f$-values summarized in Table
\ref{tab:final}.

\begin{figure}
\epsscale{1.0}
\plotone{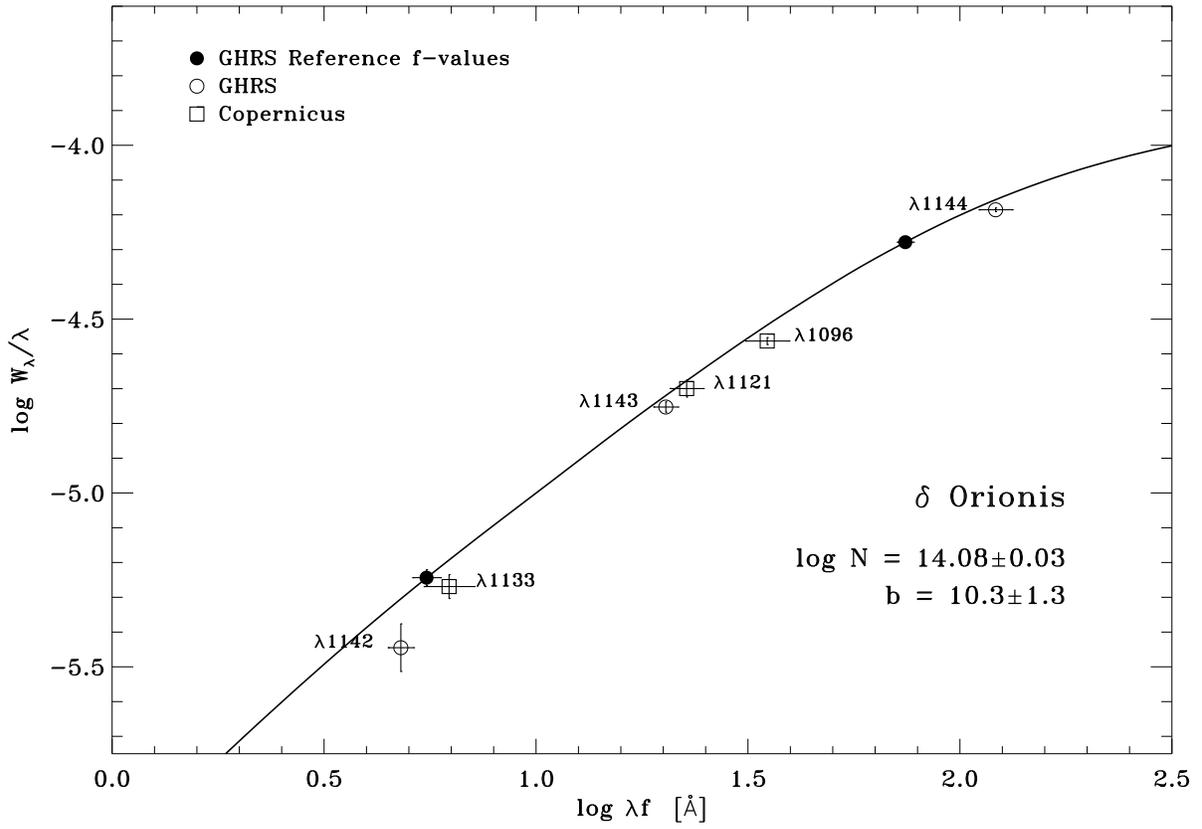}
\caption{As Figure \protect\ref{fig:mucolcog}, but for the sightline 
towards $\delta$ Ori.  The best fit column density is $\log
N(\mbox{\ion{Fe}{2}}) = 14.08\pm0.03$.
\label{fig:deloricog}}
\end{figure}

Towards \delori\ our best fit curve of growth gives $\log
N(\mbox{\FeII}) = 14.08\pm0.03$ and $b=10.3\pm1.2$ \kms; this fit is
shown in Figure \ref{fig:deloricog}.  We have not used the 2600 \AA\
transition in constructing this fit given its very large optical depth
(see Jenkins 1987).  For comparison we derive $\log N_a(\mbox{\FeII})
= 14.08\pm0.02$ by a straight integration of the $N_a(v)$ profile of
\wave{2260}.  The peak apparent optical depth of this line is $\tau_a
= 0.113\pm0.002$.  An integration of the \wave{2374} profile, which
has a peak apparent optical depth of $\tau_a = 1.440\pm0.004$, yields
$\log N_a(\mbox{\FeII}) = 14.02\pm0.01$.  Although this suggests a
modest degree of saturation in the \wave{2374} profile (all in the
component centered near $v_{\rm LSR} \approx +10$ \kms), this line is
a factor of 15 stronger than the 2260 \AA\ transition.  Given the low
optical depth of the 2260 \AA\ absorption and the relatively weak
saturation for the much stronger 2374 \AA\ line, we believe the
apparent column density derived from the \wave{2260} profile is an
accurate measure of the true column density, which is in agreement
with that derived from our curve of growth fit.

	\subsection{Summary of GHRS/\copernicus\ Oscillator Strength
	Determinations} \label{subsec:ghrssummary}

Table \ref{tab:ghrs} summarizes the results of our $f$-value
determinations using GHRS and \copernicus\ measurements.  For each
transition studied, this table gives the theoretically calculated
$f$-value from RU98 and the unweighted mean, $\langle f_\lambda
\rangle$, of our individual $f$-value determinations.  Two measures of
the errors in the mean are given: the formal statistical error and,
where appropriate, the standard deviation of the individual
measurements, $f_\lambda^i$, about the mean.  Each of the individual
$f$-value measurements is listed, along with the sightline,
instrument, and method used to derive the individual measurement.

The oscillator strengths for the transitions at 1144.938, 1143.226,
1133.665, and 1125.448 \AA\ all not only show good agreement with the
theoretical values of RU98, but also have a reasonable number of
consistent individual measurements.  The determination of the
\wave{1144} oscillator strength could potentially be prone to 
systematic errors in the shape of the true curves of growth of the
individual sightlines.  As mentioned above, it is possible that the
complicated, multi-component nature of the absorption along a given
line of sight could cause the true curve of growth to deviate from the
theoretical single-component Maxwellian curve used in our fits (see,
e.g., Morton \& Bhavsar 1979).  Towards \delori, for example,
\wave{1144} is the strongest transition placed on the curve of growth
and is thus somewhat sensitive to the adopted $b$-value, which is not
terribly well-constrained by the weaker transitions.  Towards \mucol\
and \zoph\ (to a somewhat lesser extent) this transition is
well-bracketted by secure measurements of the 1608.451 and 2586.650
\AA\ lines.  The agreement in the $f$-value derived for \wave{1144} towards  
\delori\  with those derived for the other two sightlines is
encouraging, suggesting that the systematics in the determination are
within the range of the errors.  It is also encouraging to note that
the apparent optical depth, component fitting, and curve of growth
fitting results for \mucol\ are all in good agreement.

The 1055.262, 1063.972, and 1142.366 \AA\ transitions have
formally-calculated $f$-values that are consistent with the RU98
results.  However, the individual measurements are sparse and often in
poor agreement with one another.  The oscillator strengths derived
here for the 1121.975 and 1127.098 transitions are in poor agreement
with the RU98 calculations, though there are a limited number of
individual measurements for the latter line.  In the next section we
use \fuse\ data to constrain the oscillator strengths of several of
these lines.


\section{{\em FUSE} Measurements of \ion{Fe}{2} Oscillator 
	Strengths at Far Ultraviolet Wavelengths} \label{sec:fuse}

We have derived oscillator strengths for several of the FUV \FeII\
transitions in the \fuse\ bandpass in the previous section.  The
individual determinations of the $f$-values of \FeII\
\twowave{1125.448}, 1133.665, 1143.226, and 1144.938 are in good 
agreement with one another.  Furthermore, the oscillator strengths
derived for these transitions are consistent with the theoretical
calculations of RU98.  We believe that the $f$-values for these FUV
transitions, as given in Table \ref{tab:ghrs}, are secure, and we will
use these oscillator strengths to derive estimates for the $f$-values
of \FeII\ \twowave{1112.048}, 1121.975, 1127.098, and 1142.366 using a
suite of \fuse\ observations of absorption along extended pathlengths
through the Milky Way.

The \fuse\ data used in this section, which are summarized in Table
\ref{tab:fuselog}, were taken in late-1999/early-2000.  Most of the stars 
in the current sample are located in the Large Magellanic Cloud (LMC),
with two objects in the Small Magellanic Cloud (SMC), and one in the
Milky Way.  We describe the details of the observations and reduction
of the \fuse\ datasets in \S \ref{subsec:fusereductions}.  The use of
these \fuse\ data to constrain the remaining \FeII\ oscillator
strengths is described in \S \ref{subsec:fusecog}.

	\subsection{{\em FUSE} Observations and Data Reduction}
	\label{subsec:fusereductions}

The FUSE data for this investigation were obtained during the
commissioning and early science operations stages of the FUSE mission.
An observation log can be found in Table \ref{tab:fuselog}.  Each
observation was performed with the source centered in the
$30\arcsec\times30\arcsec$ aperture of the LiF1 spectrograph
channel.\footnote{The K1-16 data were obtained as part of in-orbit
checkout activities that required multiple exposures with the object
at different locations within the aperture.  The processing of this
observation required special care to ensure that the individual
exposures were shifted and summed properly.  This will be discussed in
more detail by Kruk et al. (2000, in prep.).}  In many cases, the LiF2
channel was co-aligned well enough with the LiF1 channel that a second
set of spectra covering the 1000--1187 \AA\ wavelength range was
obtained.  Exposure times ranged from 3 ksec to 27 ksec for the
Magellanic Cloud stars observed. The time-tagged photon event lists
were processed through the standard FUSE calibration pipeline ({\tt
CALFUSE}) available at the Johns Hopkins University.  Pipeline version
1.5.3 was used for the 1999 observations, and version 1.6.8 was used
for the February 2000 observations.  The photon lists were screened
for valid data with constraints imposed for earth limb angle avoidance
and passage through the South Atlantic Anomaly.  Corrections for
detector backgrounds, Doppler shifts caused by spacecraft orbital
motions, and geometrical distortions were applied (Sahnow et al. 2000;
see also Blair et al. 2000).  No corrections were made for optical
astigmatism aberrations since the data were obtained prior to
completion of in-orbit focusing activities.  No flatfield solutions
existed at the time of these observations; therefore, we compared the
LiF1 and LiF2 spectra whenever possible to determine the significance
of the observed features.

The processed data have a nominal spectral resolution of 20--25 \kms\
(FWHM), with a relative wavelength dispersion solution accuracy of
$\sim 6$ \kms\ ($1\sigma$).  The zero point of the wavelength scale
for each individual observation is poorly determined, although the
absolute velocities of the material are not important for our
purposes.  When necessary, small velocity shifts were applied to
individual spectral lines to ensure that the line strength comparisons
were judged over a common velocity scale.  In particular, the lines at
1142, 1143, and 1144 \AA\ were systematically shifted by $-20$ \kms\
with respect to those at 1112, 1121, and 1125 \AA\ using the
dispersion solution applied to the data used in this work.

\begin{figure}
\epsscale{0.95}
\plotone{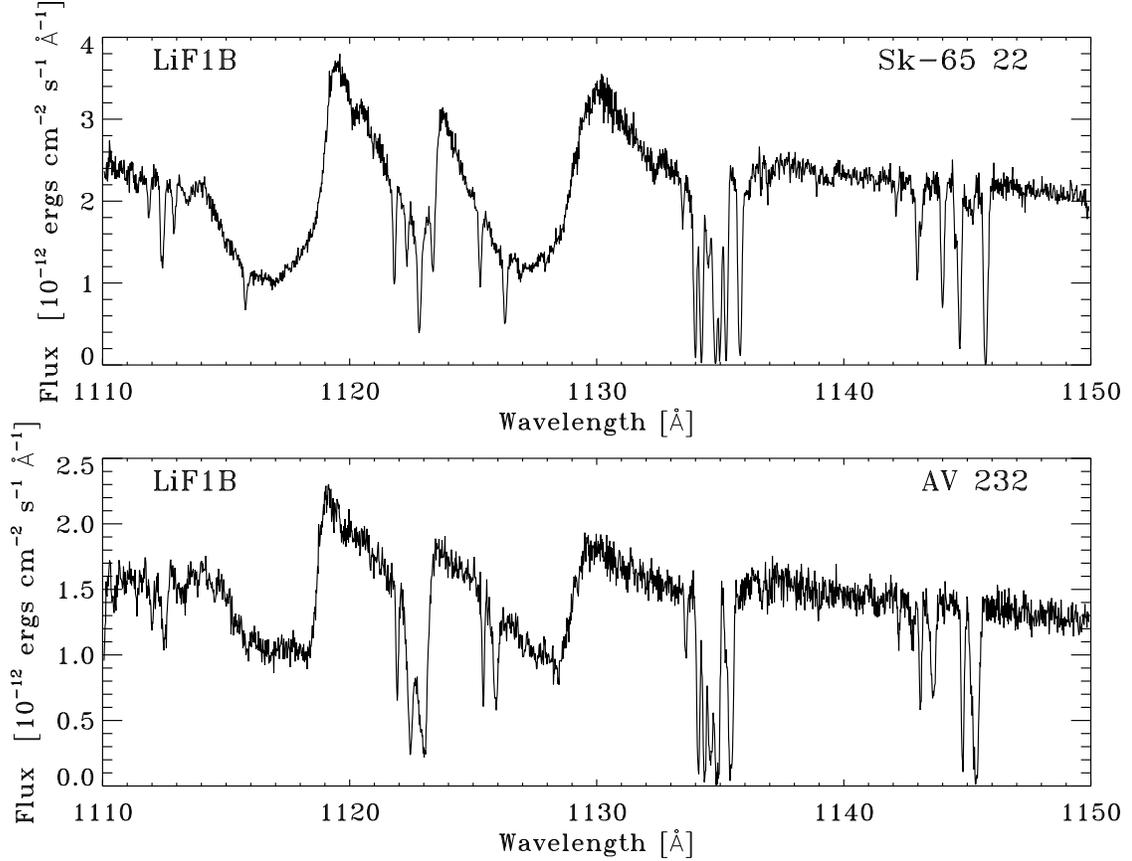}
\caption{{\em FUSE} spectra of the O6Iaf+ star Sk-65 22 
in the LMC ({\em top}) and the O7Iaf+ star AV 232 in the SMC ({\em
bottom}) over the range $1100 \la \lambda \la 1150$ \AA.  These plots
show only data from the LiF1 channel, and the data have a resolution
$\Delta v \ga 20$ km s$^{-1}$.  These spectra show absorption from
interstellar \protect\ion{Fe}{2} in the Milky Way as well as both
Magellanic Clouds at rest wavelengths of 1112.048, 1121.975, 1125.448,
1127.098, 1133.665, 1142.366, 1143.226, and 1144.938 \protect\AA.  Also seen
are interstellar lines of molecular hydrogen near 1112 \protect\AA,
\protect\ion{Fe}{3} at 1122.524 \protect\AA, and a triplet of
\protect\ion{N}{1} lines near 1135 \protect\AA.  Several prominent 
P Cygni stellar wind profiles are seen in this band, including the
\protect\ion{P}{5} doublet at 1117.977 and 1128.008 \protect\AA\ 
and the \protect\ion{Si}{4} lines at 1122.486 and 1128.339 
\protect\AA.
\label{fig:lif1bspec}}
\end{figure}

Figure \ref{fig:lif1bspec} shows portions of the \fuse\ spectra of two
O-type stars: Sk-65 22 in the LMC, and AV 232 in the SMC.  These
spectra represent only data from the LiF1 channel for each star.  Most
of the narrow interstellar lines in this bandpass are due to \FeII\
(see Table \ref{tab:final}).  Also present in these spectra are
interstellar absorption from the Lyman (0-0) band of molecular
hydrogen (near 1110.1, 1112.5, and 1115.5 \AA), from \ion{Fe}{3} at
1122.5 \AA, and from a triplet of \NI\ lines between 1134 and 1135
\AA.  These species are present in the Milky Way and the host galaxies
of these stars, in some cases causing quite complex blending of the
absorption.  Prominent stellar wind lines are also visible in this
spectrum, particularly the strong P Cygni profiles of \ion{P}{5}
\twowave{1118.0 and 1128.0} and \ion{Si}{4}
\twowave{1122.5 and 1128.3}.

We have measured equivalent widths of the \FeII\ lines in the spectra
of all of the stars listed in Table \ref{tab:fuselog} in the same
manner as described above for the GHRS data.  The measured equivalent
widths are summarized in Table \ref{tab:fuseeqw}.  For gas arising in
the LMC or SMC, many of the \FeII\ lines with well-determined
$f$-values are blended or in regions of uncertain continuum placement.
Hence, we only present the equivalent widths of absorption lines
arising from gas in the Milky Way.

The values in Table \ref{tab:fuseeqw} are averages of the equivalent
widths measured in the LiF1B and LiF2A detector segments (where the
latter were available).  When comparing spectra taken through the LiF1
and LiF2 channels, there are occasionally differences in the profiles
of absorption lines in the two channels.  This seems to be associated
with the greater degree of fixed pattern noise in the LiF2A channel.
Typically these profile differences result in large discrepancies in
the measured equivalent widths between the two channels.  In cases
where the LiF1 and LiF2 measurements differ by more than $2\sigma$, we
have discarded the measurements from both channels.

	\subsection{Curve of Growth Fitting Analysis}
	\label{subsec:fusecog}

Our determination of oscillator strengths using \fuse\ data relies on
curve of growth fitting methods and follows the procedure discussed in
\S \ref{subsec:cog}.  We adopt the mean $f$-values of \FeII\ 
\twowave{1125.448}, 1133.665, 1143.226, and 1144.938 from Table 
\ref{tab:final} as our reference oscillator strengths and use these 
to derive the $f$-values of \twowave{1112.048}, 1121.975, 1127.098,
and 1142.366.  The results of this fitting procedure are summarized in
Table \ref{tab:fuse}.  This table gives the individual $f$-value
measurements, $f_\lambda^i$, for each sightline.  Given at the bottom
of the table are the average, $\langle f_\lambda \rangle$, of our
$f$-value determinations and the RU98 theoretical oscillator strengths
for these transitions.  The average \twowave{1121}, 1127, and 1142
$f$-values were calculated using the \fuse\ determinations and the
GHRS/\copernicus\ measurements given in Table \ref{tab:ghrs}.  Also
given with the average $f$-values are the statistical error (in
parentheses) and standard deviation of the individual measurements
about the mean (in brackets).

\begin{figure}
\epsscale{1.0}
\plotone{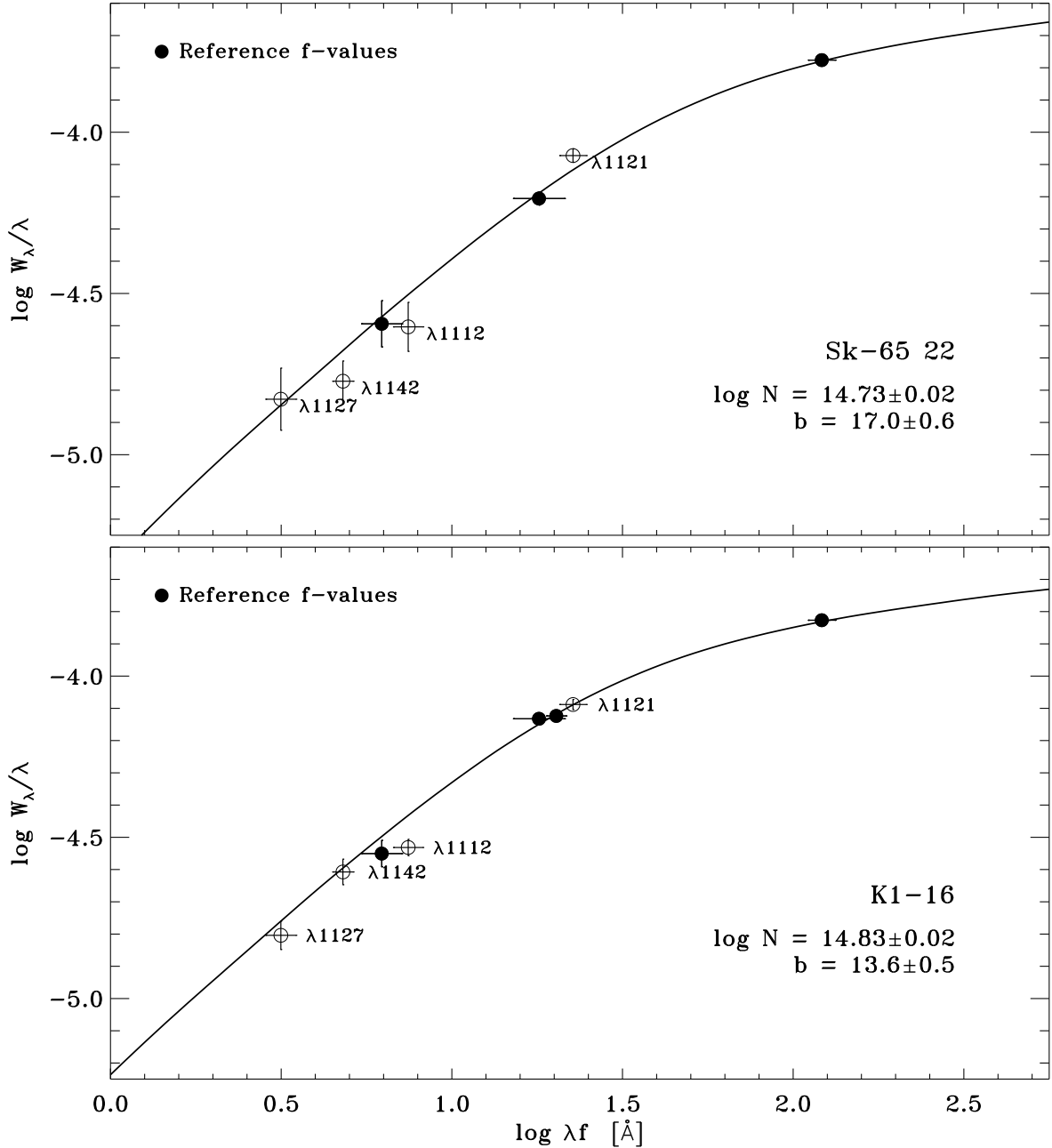}
\caption{The best-fit, single-component curves of growth  
for the Sk-65 22 ({\em top}) and K1-16 ({\em bottom})
\ion{Fe}{2} data.  The curves were fit to the  
$\lambda \lambda 1125$, 1133, 1143, and 1144 equivalent widths
(plotted as filled circles), where measured, using RU98 the oscillator
strengths.  The open circles denote FUV transitions for which we have
used the curves of growth to calculate the $f$-values.  These
transitions, at 1112, 1121, 1127, and 1142 \protect\AA, are placed on
the plots using our final oscillator strength estimates (see Table
\protect\ref{tab:final}).  The column densities and $b$-values derived from
these curves of growth are shown for each sightline.  Towards Sk-65 22
these data sample only Milky Way material.
\label{fig:fusecog}}
\end{figure}

Two sample curves of growth are shown in Figure \ref{fig:fusecog}.
These are towards the central star of a planetary nebula, K1-16, in
the Milky Way and Sk-65 22 in the LMC.  Each curve of growth
represents only Galactic absorption.  The best-fit column densities
and $b$-values are shown on the plots.  For the sightline towards
Sk-65 22, the \wave{1143} absorption is not available due to blending
with LMC \wave{1142} absorption.

\begin{figure}
\epsscale{0.8}
\plotone{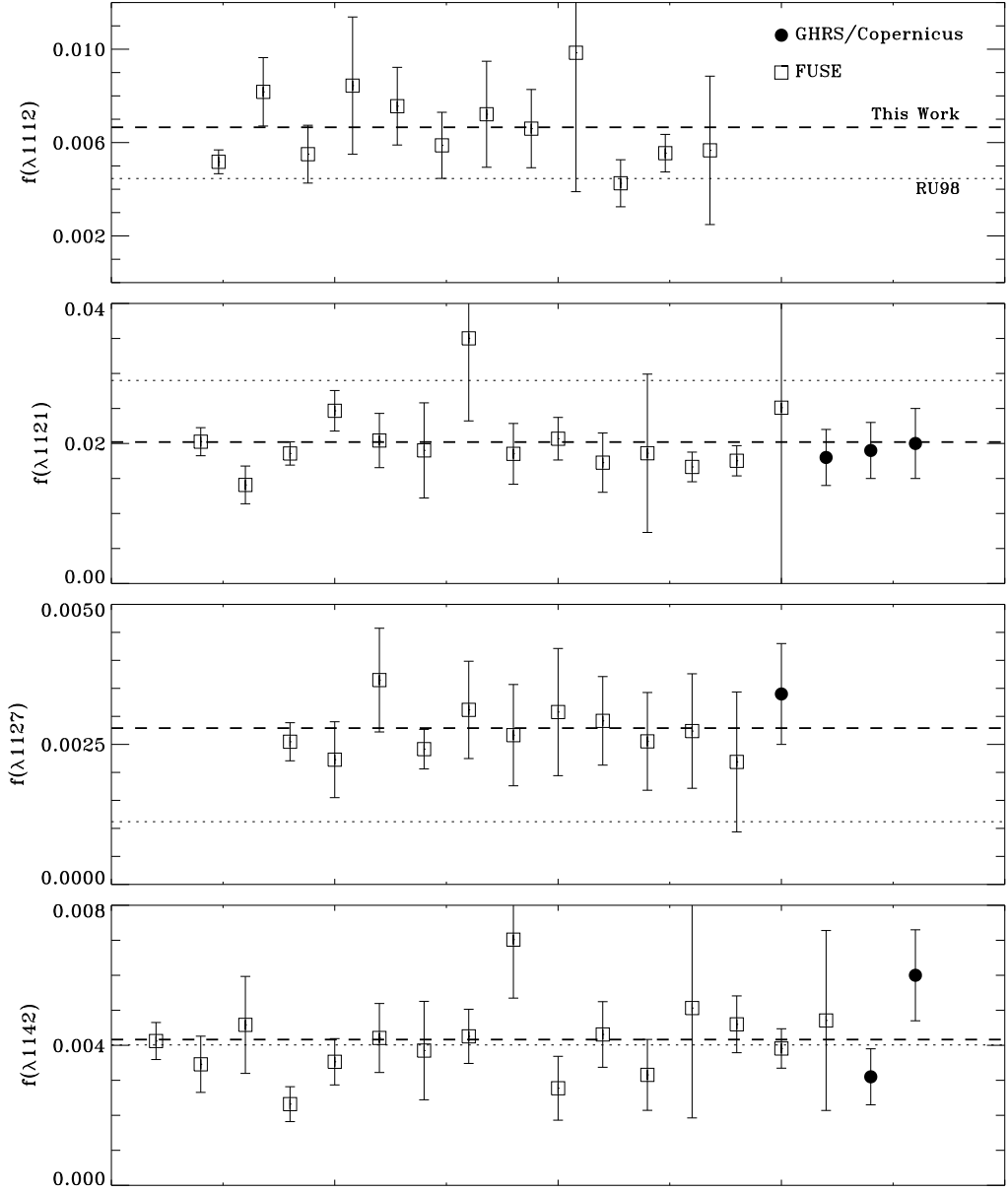}
\caption{Distribution of the individual measurements $f_\lambda^i$ of
the oscillator strengths of $\lambda \lambda1112.048$, 1121.975,
1127.098, and 1142.366.  The open squares represent estimates derived
from {\em FUSE} data, while the filled circles denote estimates
derived from GHRS and/or {\em Copernicus} data.  The dotted lines show
the oscillator strengths calculated by RU98, while the thick dashed
lines show the final average values derived in this work.
\label{fig:fdist}}
\end{figure}

Figure \ref{fig:fdist} shows the individual $f$-value determinations
and error estimates for the 1112.048, 1121.975, 1127.098, and 1142.366
\AA\ transitions.  The ordinate of each panel shows the derived
oscillator strength of the transition.  Each measurement is offset
along an otherwise meaningless abscissa.  The open squares in Figure
\ref{fig:fdist} mark measurements made using \fuse\ data (Table
\ref{tab:fuse}), while the filled circles mark estimates derived from
GHRS and/or \copernicus\ data (Table \ref{tab:ghrs}).  The dotted
lines show the oscillator strengths calculated by RU98, while the
thick dashed lines show the unweighted means of the individual
$f$-value determinations derived in this work.

Our measurements of the $f$-value for \wave{1142} are consistent with
the RU98 calculations.  The empirically-derived oscillator strengths
of the remaining three lines, however, differ significantly from the
theoretical values.  Of these, the 1121.975 \AA\ transition is the
most prone to systematic uncertainties in the curve of growth fitting
technique.  Its strength places it between the 1125 and 1144
\AA\ transitions.  It is somewhat sensitive to the adopted $b$-value,
which is determined in large part by the strong 1144 \AA\ transition.
This could, for sightlines whose curves of growth depart strongly from
the single-component, Doppler-broadened model assumed here,
potentially cause systematic uncertainties in our $f$-value
determination.  Furthermore, because we only measure one line stronger
than the 1121 \AA\ transition, the determination of $b$-values could
be skewed by systematics in the measurement of the strong 1144 \AA\
line.  The good agreement of the three GHRS/\copernicus\
determinations of $f_{\lambda 1121}$, which are quite secure, with the
majority of the \fuse\ determinations suggests these systematics play
a minor role.  Systematics of this sort are likely of order $\la 5\%$
given the the agreement of the \fuse\ and GHRS/\copernicus\
determinations.

The systematic uncertainties are likely small for the weaker 1112,
1127, and 1142 \AA\ transitions, which all fall on or near the linear
part of the curve of growth for the sightlines studied.  In these
cases the systematics associated with measurements of weak
interstellar lines in \fuse\ data are more important than the
peculiarities of the fitting procedures.  It should be noted that we
have found no systematic differences across our sample of sightlines
between the equivalent widths measured in the LiF1 and LiF2 channels
for the weak 1112, 1127, and 1142 \AA\ transitions.


\section{Discussion and Summary}
\label{sec:summary}

\begin{figure}
\epsscale{1.0}
\plotone{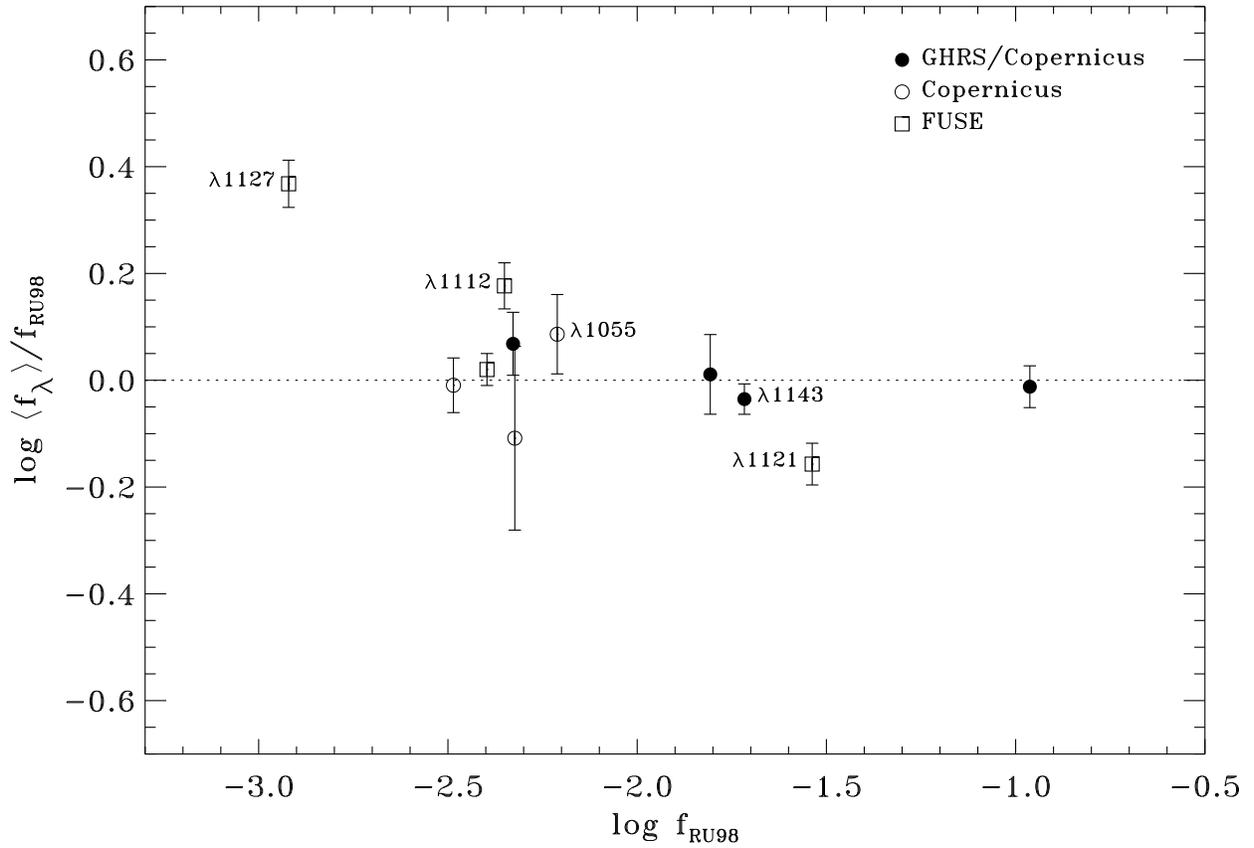}
\caption{The  ratio of the final average oscillator strengths derived 
in this work (see Table \protect\ref{tab:final}) to those calculated
by RU98 versus the RU98 strengths.  Filled circles denote measurements
that rely on {\em Copernicus} and/or GHRS data; the open circles are
base only upon {\em Copernicus} data; and the open squares indicate
estimates that include data from {\em FUSE}.  In general the agreement
between our best empirically-derived $f$-values and those calculated
by RU98.  Important exceptions include the lines at 1112.048,
1121.975, and 1127.098 \protect\AA.
\label{fig:fratio}}
\end{figure}

We have presented a self-consistent set of empirically-derived
$f$-values for the \FeII\ transitions in the range $1100 \la \lambda
\la 1150$ \AA, which are placed on an absolute scale by reference to
the NUV laboratory measurements of Mullman et al. (1997) and Bergeson
et al. (1994, 1996).  The results of our work are summarized in Table
\ref{tab:final}, where we compare the empirically-derived $f$-values
from this work with previous theoretical and empirical determinations.
Our derived oscillator strengths agree with the orthogonal operator
calculations of Raassen \& Uylings (1998), with the exception of those
at 1112.048, 1121.975, and 1127.098 \AA.  This is shown graphically in
Figure \ref{fig:fratio}, which displays the ratio of our best
$f$-values to the theoretical $f$-values of RU98 versus their values.
Over a factor of 30 range in oscillator strengths we find good
agreement (within $1.5\sigma$) between the RU98 strengths and those
derived here for 8 of the 11 lines studied in this work.

Our $f$-value determinations are in good agreement, for the most part,
with the empirical results of Lugger et al. (1982) and Shull, van
Steenberg, \& Seab (1983).  The empirical determinations by van Buren
(1986) are at times in disagreement with our values.  To some extent
the agreement of our work with these empirical studies is expected
since these studies used curve of growth fitting methods (which carry
the heaviest weight in our averages) for some of the same sightlines
used here.

We note that our derived $f$-value for \wave{1142}, $f_{\lambda 1142}
= 0.0042(3)$, is in disagreement with that of Cardelli \& Savage
(1995), who derive $f_{\lambda 1142} = 0.00247(32)$ using \ghrs\
observations of the star $\beta^1$ Scorpii.  We believe our
determination is more robust since it does not rely on a single
measurement, although the dispersion of our measurements about the
mean is relatively large for this transition.  Welty et al. (1999b;
their Appendix) discuss the potential discrepancies in FUV \FeII\
oscillator strengths, noting that the Cardelli \& Savage (1995)
$f$-value for \wave{1142} might imply the need to revise the
$f$-values of the 1143 and 1144 \AA\ lines of the same multiplet.
However, the RU98 calculations are in agreement with our empirical
determinations for the FUV $y\, ^6F^o$ multiplet transitions out of
the ground state.

As the \fuse\ mission proceeds, there may be other opportunities (and
requirements) for studying the atomic oscillator strengths at far
ultraviolet wavelengths.  Some of the more important species for which
it will be important to test the quality of the current oscillator
strengths include \ion{O}{1}, \ion{N}{1}, and the shorter wavelength
\FeII\ transitions.  These species will be well-observed by \fuse\ and
will allow researchers to study the gas-phase abundances in a variety
of environments, but only if the oscillator strengths are reasonably
well known.








\acknowledgements

We thank E. Fitzpatrick for sharing his component fitting software
with us.  This work is based on data obtained for the Guaranteed Time
Team by the NASA-CNES-CSA FUSE mission operated by the Johns Hopkins
University. Financial support to U. S. participants has been provided
by NASA contract NAS5-32985.


\begin{landscape}

\renewcommand{\mucol}{$\mu$ Col}
\renewcommand{\zoph}{$\zeta$ Oph}
\renewcommand{\delori}{$\delta$ Ori}
\renewcommand{\copernicus}{{\em Copernicus}}

\begin{deluxetable}{lllcclllllll}
\tablenum{1}
\tablecolumns{12}
\tablewidth{0pt}
\tablecaption{Recommended \ion{Fe}{2} Oscillator Strengths 
	for the {\em FUSE} Bandpass \label{tab:final}}
\tablehead{
\colhead{} & \colhead{} & \colhead{} & \colhead{} & 
\colhead{} & \colhead{} & &
\multicolumn{5}{c}{Literature $f_\lambda$-values\tablenotemark{f}} \\
\cline{8-12}
\colhead{$\lambda_c$\tablenotemark{a}} & 
\multicolumn{2}{c}{Upper Term\tablenotemark{b}} & 
\colhead{No.\tablenotemark{c}} &
\colhead{Instr.\tablenotemark{d}} & 
\colhead{$f_\lambda$\tablenotemark{e}} &  &
\colhead{RU98} & 
\colhead{Kurucz} &
\colhead{Lugger et al. } &
\colhead{van Buren} &
\colhead{Shull et al.}}
\startdata
1055.262  & $3d^5 4s 4p$ & $^6P^o_{7/2}$ &
	2  & C   & 0.0075(14) &  &  0.00615  &  0.0108 & 0.0074(10) & 
	0.0092(9) &  0.0080(12) \\
1063.972  &  $3d^5 4s 4p$ & $^6D^o_{7/2}$ &
	1  & C   &  0.0037(18) & &  0.00475  & 0.00373 & \nodata   & 
	0.0041(6) &  0.0045(9) \\
1096.877  & $3d^6 5p$ & $^6P^o_{7/2}$ &
	3  & C   & 0.032(4)  &  &  0.0327  & 0.0565 & 0.032(5)   & 
	0.0370(15) &  0.032(5) \\ 
1112.048  & $3d^6 5p$ & $^6F^o_{11/2}$ &
	12 & F   & 0.0062(9)  &  &  0.00446  & 0.00826 & \nodata   & 
	\nodata   &  \nodata \\
1121.975  & $3d^5 4s 4p$ & $^6P^o_{7/2}$ &
	17 & CGF &  0.0202(20) &  &  0.0290   & 0.0189 & 0.0203(12) & 
	0.0227(16) &  0.020(3) \\
1125.448  & $3d^6 5p$ & $^6D^o_{7/2}$ &
	2  & CG  &  0.016(3)   & &  0.0156   & 0.0261 & \nodata   & 
	0.0205(16) &  0.011(3) \\
1127.098  & $3d^6 5p$ & $^6D^o_{9/2}$ &
	12 & CGF &  0.0028(3)  & &  0.00112  & 0.00228 & \nodata   & 
	\nodata   &  \nodata \\
1133.665  & $3d^5 4s 4p$ & $^6D^o_{7/2}$ &
	3  & CG  &  0.0055(8)  & &  0.0047   & 0.0125 & 0.0048(5)  & 
	0.0074(11) &  0.0060(9) \\
1142.366\tablenotemark{g} & $3d^5 4s 4p$ & $y\, ^6F^o_{7/2}$ &
	18 & CGF &  0.0042(3)  & &  0.00401  & 0.00573 & 0.0050(5)  & 
	0.0060(7)  &   0.0050(15)\tablenotemark{h} \\
1143.226  & $3d^5 4s 4p$ & $y\, ^6F^o_{9/2}$ &
	5  & CG  &  0.0177(12) & &  0.0192   & 0.0268 & 0.0133(7)  & 
	0.0200(18) & 0.013(4)  \\
1144.938  & $3d^5 4s 4p$ & $y\, ^6F^o_{11/2}$ &
	5  & G   &  0.106(10)  & &  0.1090   & 0.1122 & 0.1050(5)  & 
	0.126(9)  &   0.105(21) \\
\enddata
\tablenotetext{a}{Vaccum wavelengths (in \AA) from Morton (2000).} 
\tablenotetext{b}{Designation of the upper term of the transition.  All
	transitions arise from the ground state, $3d^6 (a\, ^5D) 4s
	^6D_{9/2}$.}
\tablenotetext{c}{Number of individual $f_\lambda^i$ determinations used 
	in deriving the final value presented in this table.}
\tablenotetext{d}{The instrument(s) used in deriving each $f_\lambda$, 
	where C, G, and F denote {\em Copernicus}, GHRS, and {\em
	FUSE}, respectively.}
\tablenotetext{e}{Final derived oscillator stengths from the current
	work.  The numbers in parentheses denote the uncertainties in
	the last digit(s).}
\tablenotetext{f}{Selected literature values of \ion{Fe}{2} oscillator 
	strengths for comparison with those derived in the current
	work.  The columns give literature $f$-values from the
	calculations of Raassen \& Uylings (1998; adopted by Morton
	2000) and Kurucz (1988), and from the empirical
	(astrophysical) determinations by Lugger et al. (1982), van
	Buren (1986), and Shull et al. (1983), respectively.  The
	numbers in parentheses denote the uncertainties in the last
	digit(s).}
\tablenotetext{g}{Cardelli \& Savage (1995) have also used the GHRS 
	to empirically estimate this $f$-value.  They derive
	$f_{\lambda 1142}=0.00247(32)$ based only on the sightline
	towards $\beta^1$ Scorpii using a curve-of-growth fit.}
\tablenotetext{h}{Shull et al. (1983) give a value of 0.050(15) in their
	Table 3.  This is a typo that has been corrected for the
	current table (J.M. Shull, 2000, priv. comm.).}
\end{deluxetable} 

\end{landscape}


\begin{deluxetable}{rlccccc}
\tablenum{2}
\tablecolumns{7}
\tablewidth{0pc}
\tablecaption{GHRS Observation Log\tablenotemark{a}      
	\label{tab:ghrslog}}
\tablehead{
\colhead{Spectra Range} & 
\colhead{Rootname\tablenotemark{b}} &
\colhead{Exp. Time} & 
\colhead{Grating} & 
\colhead{Resolution\tablenotemark{c}} &
\colhead{Aper.\tablenotemark{d}} & 
\colhead{COSTAR?\tablenotemark{e}} \\
\colhead{[\AA]} & \colhead{} &\colhead{[sec]} & 
\colhead{} & \colhead{[km s$^{-1}$]} & \colhead{} & \colhead{}
}
\startdata
\cutinhead{$\mu$ Columbae}
1116-1124
        & Z2X30108T & 81.6 & G140M & 16. & LSA & Yes \\
        & Z2X30109T & 81.6 & G140M & 16. & LSA & Yes \\
1131-1137 
        & Z2C0011AP & 108.8  & Ech-A & 3.5 & LSA & Yes \\
1142-1148
        & Z2AF011AT & 108.8 & Ech-A & 3.5 & LSA & Yes \\
        & Z2C0030RT & 108.8 & Ech-A & 3.5 & LSA & Yes \\
        & Z308010OT & 108.8 & Ech-A & 3.5 & LSA & Yes \\
        & Z3JN010OT & 108.8 & Ech-A & 3.5 & LSA & Yes \\
\cutinhead{$\delta$ Orionis A}
1127-1164
	& Z1850107T & 217.6 & G160M & 21. & SSA & No \\
2256-2265
	& Z185020KT & 13.6  & Ech-B & 3.5 & SSA & No \\
	& Z185020LT & 13.6  & Ech-B & 3.5 & SSA & No \\
	& Z185020MT & 13.6  & Ech-B & 3.5 & SSA & No \\

2367-2377
	& Z185020NT & 13.6  & Ech-B & 3.5 & SSA & No \\
	& Z185020OT & 13.6  & Ech-B & 3.5 & SSA & No \\
	& Z185020PT & 13.6  & Ech-B & 3.5 & SSA & No \\

2599-2609
	& Z185020QT & 13.6  & Ech-B & 3.5 & SSA & No \\
	& Z185020RT & 13.6  & Ech-B & 3.5 & SSA & No \\
	& Z185020ST & 13.6  & Ech-B & 3.5 & SSA & No \\

\cutinhead{$\zeta$ Ophiuchi}
1126-1154 
	& Z0HU020WM & 68.0 & G140M & 15. & SSA & No \\
	& Z0HU520UT & 68.0 & G140M & 15. & SSA & No \\
1603-1638
	& Z0HU012AT & 13.6 & G160M & 14. & LSA & No \\
2241-2279
	& Z0HU013JM & 13.6 & G200M & 11. & LSA & No \\
2334-2347
	& Z0XZ010BT & 5.6 & Ech-B  & 3.5 & SSA & No \\
\enddata
\tablenotetext{a}{This table describes the GHRS exposures used 
	in this work. For $\mu$ Col and $\zeta$ Oph, we list only
	those datasets covering $\lambda < 1200$ \AA.  The longer
	wavelength GHRS data for $\mu$ Col are described by Howk et
	al. (1999) and the $\zeta$ Oph data are discussed by Savage et
	al. (1992).}
\tablenotetext{b}{STScI archival rootname.}
\tablenotetext{c}{Approximate velocity resolution (FWHM) in km s$^{-1}$.}
\tablenotetext{d}{The entrance aperture used for the observation.  Here
	SSA refers to the small science aperture (pre-COSTAR:
	$0\farcs24\times0\farcs24$; post-COSTAR:
	$0\farcs22\times0\farcs22$) and LSA refers to the large
	science aperture (pre-COSTAR: $2\farcs00\times2\farcs00$;
	post-COSTAR: $1\farcs74\times1\farcs74$)}
\tablenotetext{e}{Denotes observations taken before (No) or after (Yes)
	the installation of the Corrective Optics Space Telescope
	Axial Replacement (COSTAR) unit.  }
\end{deluxetable}


\newcommand{\cop}{{\em Cop}}

\begin{deluxetable}{lccccccccc}
\tablenum{3}
\tablecolumns{9}
\tablewidth{0pc}
\tablecaption{GHRS/{\em Copernicus} \ion{Fe}{2} Equivalent Width 
	Measurements \label{tab:ghrseqw}}
\tablehead{
\colhead{$\lambda_c$\tablenotemark{a}} & 
\multicolumn{2}{c}{$\mu$ Col\tablenotemark{b}} & &
\multicolumn{2}{c}{$\delta$ Ori\tablenotemark{c}} & &
\multicolumn{2}{c}{$\zeta$ Oph\tablenotemark{d}} \\
\cline{2-3} \cline{5-6} \cline{8-9}
\colhead{[\AA]} & 
\colhead{$W_\lambda$ [m\AA]} & \colhead{Instr.\tablenotemark{e}} & &
\colhead{$W_\lambda$ [m\AA]} & \colhead{Instr.\tablenotemark{e}} & &
\colhead{$W_\lambda$ [m\AA]} & \colhead{Instr.\tablenotemark{e}} }
\startdata
1055.262 & $11.2\pm2.0$ & \cop & 
	& \nodata & \nodata & 
	& $21(\pm3)$ & \cop \\
1063.972 & $7\pm3$      & \cop &
	& \nodata & \nodata & 
	& \nodata & \nodata \\
1096.877 & $53\pm3$     & \cop &
	& $30.0\pm0.7$ & \cop &
	& $52(\pm3)$ & \cop \\
1121.975 & $34\pm4$     & G140M &
	& $22.4\pm1.3$ & \cop &
	& $42(\pm3)$ & \cop \\
1125.448 & $27\pm4$     & G140M &
	& \nodata & \nodata & 
	& $39(\pm3)$ & \cop \\
1127.098 & \nodata  &  \nodata &
	& \nodata & \nodata & 
	& $10.5\pm2.1$ & \cop/G140M \\
1133.665 & $14.0\pm2.5$ & Ech-A &
	& $6.1\pm0.5$ & \cop &
	& $15.5\pm1.9$ & \cop/G140M \\
1142.366 & \nodata  &  \nodata & 
	& $4.1\pm0.7$ & G160M &
	& $17.9\pm1.9$ & \cop/G140M \\
1143.226 & $34.8\pm1.1$ & Ech-A &
	& $20.2\pm0.7$ & G160M &
	& $37.9\pm2.2$ & \cop/G140M \\
1144.938 & $112.9\pm2.1$ & Ech-A &
	& $74.7\pm0.9$ & G160M &
	& $80.7\pm1.4$ & G140M \\
1608.451 & $144.4\pm2.4$ & Ech-A &
	& \nodata & \nodata & 
	& $103\pm3$ & G160M \\
2249.877 & $14.2\pm0.4$ & Ech-B &
	& \nodata & \nodata & 
	& $22.5\pm1.1$  & Ech-B \\
2260.780 & $21.9\pm0.6$ & Ech-B &
	& $12.9\pm0.7$ & Ech-B &
	& $31\pm4$ & G200M \\
2344.214 & $302\pm4$  & Ech-B &
	& \nodata & \nodata & 
	& $195\pm4$ & Ech-B \\
2374.461 & $182\pm3$ & Ech-B &
	& $125\pm3$ & Ech-B &
 	& $149.2\pm1.3$ & Ech-B \\
2586.650 & $291\pm4$ & Ech-B &
	& \nodata & \nodata & 
	& \nodata & \nodata \\
2600.173 & $388\pm5$  & Ech-B &
	& $370\pm7$ & Ech-B &
	& \nodata & \nodata \\
\enddata
\tablenotetext{a}{Central wavelengths from Morton (2000).}
\tablenotetext{b}{The {\em Copernicus} equivalent width measurements
	towards $\mu$ Col are from Shull \& York (1977).  The GHRS
	measurements for $\lambda_c >1133$ \AA\ are from Howk et
	al. (1999), while the remaining GHRS G140M measurements are
	from this work.}
\tablenotetext{c}{The $\delta$ Ori {\em Copernicus} measurements are
	from Bohlin et al. (1983) while the GHRS measurements are from
	this work.}
\tablenotetext{d}{The $\zeta$ Oph {\em Copernicus} measurements are 
	from Morton (1975).  The $\lambda \lambda 2249, 2374$ Ech-B
	GHRS measurements are from Savage et al. (1992).  The
	remaining GHRS measurements are from this work.  Where no
	error measurement was quoted for the {\em Copernicus}
	equivalent widths we assumed a $1 \sigma$ error of 3 m\AA.
	Such errors are given in parentheses.  Both GHRS and {\em
	Copernicus} equivalent width measurements were available for
	several transitions. In most of these cases the measurements
	agreed to within $1\sigma$ of the GHRS measurements.  Where
	two measurements are available, and the {\em Copernicus} data
	seem of high quality, we averaged the equivalent widths from
	these two instruments. }
\tablenotetext{e}{The instrument with which the measurements were made.
	In this case {\em Cop} refers to measurements made with the
	{\em Copernicus} satellite, while GHRS measurements are listed
	according to the grating used: Ech-A, Ech-B, G140M, G160M, or
	G200M.}
\end{deluxetable}


\begin{deluxetable}{llcl}
\tablenum{4}
\tablecolumns{4}
\tablewidth{0pc}
\tablecaption{Near Ultraviolet \ion{Fe}{2} Oscillator Strengths
	\label{tab:nuvfvalues}}
\tablehead{
\colhead{$\lambda_c$ [\AA]\tablenotemark{a}} &
\colhead{$f_\lambda$\tablenotemark{b}} &
\colhead{Ref.\tablenotemark{c}} &
\colhead{$f_{\rm RU98}$\tablenotemark{d}}
}
\startdata
1608.451 & 0.058(5)    & 1 & 0.054  \\
2249.877 & 0.00182(14) & 2 & 0.00218  \\
2260.780 & 0.00244(19) & 2 & 0.00262  \\
2344.214 & 0.114(2)    & 2 & 0.125  \\
2374.461 & 0.0313(14)  & 2 & 0.0329  \\
2586.650 & 0.0691(25)  & 2 & 0.0709  \\
2600.173 & 0.239(4)    & 2 & 0.242  \\
\enddata
\tablenotetext{a}{Central wavelength from Morton (2000).}
\tablenotetext{b}{Adopted experimental oscillator strength.}
\tablenotetext{c}{Reference for adopted experimental oscillator 
strengths: (1) Mullman et al. (1997); (2) Bergeson et al. (1996).}
\tablenotetext{d}{Theoretical oscillator strength from Raassen \&
Uylings (1998) for comparison.}
\end{deluxetable}


\begin{deluxetable}{lllllcc}
\tablenum{5}
\tablecolumns{7}
\tablewidth{0pc}
\tablecaption{Summary of GHRS/{\em Copernicus} Results
        \label{tab:ghrs}}
\tablehead{
\colhead{$\lambda_c$\tablenotemark{a}} & 
\colhead{$f_{\rm RU98}$\tablenotemark{b}} & 
\colhead{$\langle f_\lambda \rangle$\tablenotemark{c}} &
\colhead{$f_\lambda^i$ \tablenotemark{d}} & 
\colhead{Star} & 
\colhead{Instrument\tablenotemark{e}} & 
\colhead{Method\tablenotemark{f}}}
\startdata
1055.262 & 0.00615 & 0.0075(14)[--] & 
		0.0061(16) & \mucol & \cop & CoG \\
          & & & 0.0088(22) & \zoph  & \cop & CoG \\
1063.972 & 0.00475 & 0.0037(18)[--] & 
		0.0037(18) & \mucol & \cop & CoG \\
1096.877 & 0.0327 & 0.032(4)[3] & 
		0.034(7) & \mucol & \cop & CoG \\
	&  &  & 0.028(6) & \delori & \cop & CoG \\
	&  &  & 0.033(9) & \zoph  & \cop & CoG \\ 
1121.975 & 0.0290 & 0.019(3)[1] &
		0.018(4) & \mucol & G140M & CoG \\
	&  &  & 0.019(4) & \delori & \cop & CoG \\
	&  &  & 0.020(5) & \zoph  & \cop & CoG \\	
1125.448 & 0.0156 & 0.016(3)[3] & 
		0.014(4) & \mucol & G140M & CoG \\
	&  &  & 0.018(4) & \zoph  & \cop & CoG \\
1127.098 & 0.00112 & 0.0034(9)[--] & 
		0.0034(9) & \zoph & G140M/\cop & CoG \\
1133.665 & 0.0047  & 0.0055(8)[11] &
		0.0067(18) & \mucol & Ech-A & CoG \\
	&  &  & 0.0046(10) & \delori & \cop & CoG \\
	&  &  & 0.0052(12) & \zoph  & G140M/\cop & CoG \\
1142.366 & 0.00401 & 0.0045(8)[--] & 
		0.0031(8) & \delori & G160M & CoG \\
	&  &  & 0.0060(13) & \zoph & G140M/\cop & CoG \\
1143.226 & 0.0192 & 0.0177(12)[18] & 
		0.018(4)   & \mucol & Ech-A  & CoG  \\
	&  &  & 0.0206(8) & \mucol & Ech-A  & AOD  \\
	&  &  & 0.0181(13) & \mucol & Ech-A  & CF   \\
	&  &  & 0.016(3)   & \delori & G160M & CoG \\
	&  &  & 0.016(3)   & \zoph  & G140M/\cop & CoG \\
1144.938 & 0.1090 & 0.106(10)[11] & 
		0.11(3)   & \mucol & Ech-A & CoG \\
	&  &  & 0.107(4)  & \mucol & Ech-A & AOD \\
	&  &  & 0.120(18) & \mucol & Ech-A & CF  \\
	&  &  & 0.09(3)   & \delori & G160M & CoG \\
	&  &  & 0.104(17) & \zoph  & G140M & CoG \\
\enddata
\tablenotetext{a}{Vaccum wavelengths (in \AA) from Morton (2000).} 
\tablenotetext{b}{Theoretical values of $f_\lambda$ calculated by 
	Raassen \& Uylings (1998).  These oscillator strengths are
	adopted in the new compilation of Morton (2000). }
\tablenotetext{c}{Average values of $f_\lambda$ with $1 \sigma$ 
	statistical uncertainties of the last digits in parentheses.
	The standard deviations of the individual measurements
	$f_\lambda^i$ about the mean are given in square brackets.}
\tablenotetext{d}{Individual measurements  $f_\lambda^i$ of the 
	oscillator strengths with $1\sigma$ errors in the last digit(s)
	given in parentheses.}
\tablenotetext{e}{Instrument used in deriving the individual oscillator
	strength estimates: {\em Cop} refers to data taken with the
	{\em Copernicus} satellite, while Ech-A, G140M, and G160M
	refer to the gratings used in GHRS observations.  The {\em
	Copernicus} data for \mucol, \zoph, and \delori\ are taken
	from Shull \& York (1977), Morton (1975), and Bohlin et
	al. (1983), respectively.  }
\tablenotetext{f}{Method used for deriving the individual $f$-value
	measurement, $f_\lambda^i$: CoG - curve of growth fitting; AOD
	- apparent optical depth/column density method; and CF -
	component fitting.}
\end{deluxetable}


\begin{deluxetable}{lllcccl}
\tablenum{6}
\tablecolumns{7}
\tablewidth{0pc}
\tablecaption{{\em FUSE} Observation Log       
	\label{tab:fuselog}}
\tablehead{
\colhead{} & 
\colhead{Altern.} & 
\colhead{} & 
\colhead{} & 
\colhead{Exp. Time} & 
\colhead{S/N} & 
\colhead{} \\
\colhead{Star} & 
\colhead{Name} &
\colhead{Dataset} & 
\colhead{Date\tablenotemark{a}} &
\colhead{[ksec]} & 
\colhead{($\lambda1125$)\tablenotemark{b}} & 
\colhead{Notes}}
\startdata
K1-16\tablenotemark{c}
          & \nodata & I81103**  & 10/15/99 & 40.0  & 30 & CSPN \\
AV 232    & Sk 80     & X0200201 & 09/25/99 & 10.3 & 18 & SMC \\
HD 5980   & Sk 78     & X0240202 & 10/20/99 & 3.2  & 18 & SMC \\
Sk-65 22  & HD 270952 & P1031002 & 12/20/99 & 27.2 & 17 & LMC \\
Sk-67 104 & HD 36402  & P1031302 & 12/17/99 & 5.1  & 13 & LMC \\
Sk-68 80  & HD 36521  & P1031402 & 12/17/99 & 9.7  & 17 & LMC \\
Sk-69 246 & HD 38282  & P1031802 & 12/16/99 & 22.1 & 10 & LMC \\
Sk-67 211 & HD 269810 & P1171603 & 12/20/99 & 8.2  & 16 & LMC \\
Sk-67 69  & \nodata  & P1171703 & 12/20/99 & 6.2  & 8  & LMC \\
Sk-67 167 & LH76:21   & P1171902 & 12/17/99 & 2.8  & 11 & LMC \\ 
Sk-66 100 & \nodata  & P1172303 & 12/20/99 & 7.1  & 8  & LMC \\
Sk-70 115 & HD 270145 & P1172601 & 02/12/00 & 5.2  & 13 & LMC \\
BI 229    & \nodata  & P1172801 & 02/11/00 & 5.6  & 17 & LMC \\
Sk-67 111 & LH60:53   & P1173001 & 02/11/00 & 8.0  & 19 & LMC \\
Sk-69 249 & HD 269927 & P1174601 & 02/09/00 & 7.2  & 23 & LMC \\ 
\enddata
\tablenotetext{a}{Date of observation.}
\tablenotetext{b}{Empirical signal to noise estimate near 1125 \AA\
	in the LiF1 channel.}
\tablenotetext{c}{The K1-16 data were obtained as part 
	of in-orbit checkout activities.  Since the object was
	observed at several different locations within the aperture,
	the processing of these data required special processing to
	ensure that the individual exposures were shifted and summed
	properly.  This will be discussed in more detail by Kruk et
	al. (2000, in prep.).}
\end{deluxetable}


\begin{deluxetable}{lcccccccc}
\tablenum{7}
\tablecolumns{9}
\tablewidth{0pc}
\tablecaption{{\em FUSE} \ion{Fe}{2} Equivalent Width Measurements
        \label{tab:fuseeqw}}
\tablehead{
\colhead{} & \multicolumn{8}{c}{$W_\lambda$ [m\AA]\tablenotemark{a}} \\
\cline{2-9}
\colhead{Star} & 
\colhead{$\lambda 1112.048$} & 
\colhead{$\lambda 1121.975$} &
\colhead{$\lambda 1125.448$} &
\colhead{$\lambda 1127.098$} &
\colhead{$\lambda 1133.665$} &
\colhead{$\lambda 1142.366$} &
\colhead{$\lambda 1143.226$} &
\colhead{$\lambda 1144.938$} }
\startdata
K1-16     & $ 33\pm 2$ & $ 92\pm 3$ & $ 83\pm 3$ & $ 18\pm 2$ 
            & $ 32\pm 3$ & $ 28\pm 3$ & $ 86\pm 3$ & $171\pm 4$ \\
AV 232    & $ 45\pm 5$ & $ 65\pm 6$ & $ 63\pm 4$ & $ 16\pm 4$ 
            & $ 42\pm 5$ & $ 24\pm 4$ & $ 76\pm 6$ & $132\pm 4$ \\
HD 5980   &  \nodata  &  \nodata  & $ 64\pm 5$ & $ 20\pm 4$ 
            & $ 34\pm 5$ & $ 25\pm 6$ & $ 60\pm 5$ & $128\pm 4$ \\
Sk-67 05\tablenotemark{c}
          &  \nodata  & $ 85\pm 2$ & $ 78\pm 2$ & $ 16\pm 2$ 
            & $ 38\pm 5$ & $ 16\pm 3$ &  \nodata  & $173\pm 4$ \\
Sk-65 22  & $ 28\pm 5$ & $ 95\pm 5$ & $ 70\pm 3$ & $ 17\pm 4$ 
            & $ 29\pm 5$ & $ 19\pm 3$ &  \nodata  & $192\pm 4$ \\
Sk-67 104 &  \nodata  & $ 85\pm 6$ & $ 73\pm 4$ & $ 18\pm 5$ 
            & $ 37\pm 5$ & $ 28\pm 5$ &  \nodata  & $148\pm 5$ \\
Sk-68 80  & $ 41\pm 4$ & $ 84\pm 5$ & $ 74\pm 3$ & $ 16\pm 3$ 
            & $ 25\pm 4$ & $ 21\pm 3$ &  \nodata  & $151\pm 4$ \\
Sk-69 246 & $ 42\pm 7$ & $107\pm 6$ & $ 87\pm 7$ & $ 18\pm 4$ 
            & $ 25\pm 4$ & $ 26\pm 3$ &  \nodata  & $213\pm 6$ \\
Sk-67 211 & $ 32\pm 4$ & $104\pm 7$ & $ 71\pm 5$ & $ 16\pm 4$ 
            & $ 28\pm 5$ & $ 39\pm 4$ &  \nodata  & $145\pm10$ \\
Sk-67 69  & $ 40\pm 9$ & $ 83\pm 8$ & $ 80\pm 7$ &  \nodata  
            & $ 31\pm 4$ & $ 18\pm 5$ &  \nodata  & $172\pm 9$ \\
Sk-67 167 &  \nodata  & $ 74\pm 5$ & $ 64\pm 5$ &  \nodata  
            & $ 27\pm 4$ & $ 23\pm 4$ & $ 69\pm 4$ & $138\pm 4$ \\
Sk-66 100 & $ 40\pm 5$ & $ 73\pm 5$ & $ 72\pm 5$ &  \nodata  
            & $ 41\pm 6$ & $ 23\pm 5$ & $ 70\pm 6$ & $125\pm 5$ \\
Sk-70 115 & $ 56\pm 6$ & $ 96\pm 6$ & $ 90\pm 5$ &  \nodata  
            & $ 31\pm 6$ & $ 32\pm 5$ & $ 93\pm 5$ & \nodata \\
BI 229    & $ 19\pm 4$ & $ 62\pm 3$ & $ 55\pm 3$ & $ 13\pm 4$ 
            & $ 27\pm 5$ & $ 22\pm 3$ & $ 76\pm 5$ & $160\pm 6$ \\
Sk-67 111 & $ 29\pm 3$ & $ 70\pm 4$ & $ 64\pm 3$ &  \nodata  
            & $ 33\pm 3$ & $ 22\pm 3$ & $ 73\pm 3$ & $145\pm 3$ \\
Sk-69 249 & $ 33\pm 4$ & $ 91\pm 3$ & $ 71\pm 3$ & $ 14\pm 3$ 
            & $ 33\pm 3$ & $ 29\pm 2$ &  \nodata  & \nodata \\
\enddata
\tablenotetext{a}{The equivalent width measurements and $1\sigma$ errors 
	given here are averages of the LiF1B and LiF2A observations
	for these sightlines.}
\tablenotetext{b}{The equivalent widths for Sk-67 05 are from 
	Friedman et al. (2000).}
\end{deluxetable}


\begin{deluxetable}{llllll}
\tablenum{8}
\tablecolumns{6}
\tablewidth{0pc}
\tablecaption{Summary of {\em FUSE} $f$-value Determinations
        \label{tab:fuse}}
\tablehead{
\colhead{} & \colhead{} & 
	\multicolumn{4}{c}{$f_\lambda^i$\tablenotemark{a}} \\
\cline{3-6}
\colhead{Star} & \colhead{} & 
\colhead{$\lambda1112.048$} & \colhead{$\lambda1121.975$} & 
\colhead{$\lambda1127.098$} & \colhead{$\lambda1142.366$}}
\startdata
K1-16     & & 0.0052(5)  & 0.0203(20) & 0.0025(3) & 0.0041(5) \\ 
AV 232    & & 0.0082(15) & 0.014(3)   & 0.0022(7) & 0.0035(8) \\
HD 5980   & & \nodata   & \nodata   & 0.0036(9) & 0.0046(14) \\ 
Sk-67 05  & & \nodata   & 0.0186(17) & 0.0024(4) & 0.0023(5)  \\
Sk-65 22  & & 0.0055(12) & 0.025(3)   & 0.0031(9) & 0.0035(7)  \\
Sk-67 104 & & \nodata   & 0.020(4)   & 0.0027(9) & 0.0042(10) \\
Sk-68 80  & & 0.008(4)   & 0.019(7)   & 0.0031(11)& 0.0038(14) \\
Sk-69 246 & & 0.0076(16) & \nodata   & 0.0029(8) & 0.0043(8)  \\
Sk-67 211 & & 0.0059(14) & 0.035(12)  & 0.0026(9) & 0.0070(17) \\
Sk-67 69  & & 0.0072(23) & 0.019(4)   & \nodata  & 0.0028(9) \\
Sk-67 167 & & \nodata   & 0.021(3)   & \nodata  & 0.0043(9) \\
Sk-66 100 & & 0.0066(17) & 0.017(4)   & \nodata  & 0.0032(10) \\
Sk-70 115 & & 0.010(6)   & 0.019(11)  & \nodata  & 0.005(3)  \\
BI 229    & & 0.0043(10) & 0.0167(21) & 0.0027(10)& 0.0046(8) \\
Sk-67 111 & & 0.0055(8)  & 0.0175(22) & \nodata  & 0.0039(6)  \\
Sk-69 249 & & 0.0076(16) & \nodata   & 0.0029(8) & 0.0043(8)  \\
\\
\multicolumn{1}{c}{$\langle f_\lambda \rangle$\tablenotemark{b}}
          & & 0.0062(9)[16] & 0.0202(20)[46] & 0.0028(3)[5] 
		& 0.0042(3)[11] \\
\multicolumn{1}{c}{$f_{\rm RU98}$\tablenotemark{c}} 
	  & & 0.00446 & 0.0290  & 0.00112 & 0.00401 \\
\enddata
\tablenotetext{a}{Individual measurements  $f_\lambda^i$ of the 
	oscillator strengths with $1\sigma$ errors in the last digit(s)
	given in parentheses.}
\tablenotetext{b}{Average value of $f_\lambda$ with $1 \sigma$ 
	statistical uncertainties of the last digits in parentheses.
	The standard deviations of the individual measurements
	$f_\lambda^i$ about the mean are given in square brackets.}
\tablenotetext{c}{Theoretical values of $f_\lambda$ from the calculations 
	of Raassen \& Uylings (1998).}
\end{deluxetable}



\clearpage

\clearpage

\clearpage

\end{document}